\documentclass[aps,prb,twocolumn,superscriptaddress]{revtex4-2}
\usepackage[pdftex]{graphicx}
\usepackage{booktabs}
\usepackage{multirow}
\usepackage[mediumspace,mediumqspace,Grey,squaren]{SIunits}
\usepackage{color}
\usepackage{natbib}
\usepackage{amsmath,amssymb}
\usepackage{lipsum}
\usepackage{float}
\usepackage[normalem]{ulem}
\usepackage{bm}
\usepackage[euler]{textgreek}
\graphicspath{ {./Figures/} }

\begin{document}

%\preprint{APS/123-QED}

\title{Photocurrent-driven transient symmetry breaking in the Weyl semimetal TaAs}% Transient photoinduced symmetry breaking in the Weyl semimetal TaAs

\author{N. Sirica}
\email{nsirica@lanl.gov}
\affiliation{Center for Integrated Nanotechnologies, Los Alamos National Laboratory, Los Alamos, NM 87545, USA}
\author{P. P. Orth}
\affiliation{Ames Laboratory, Ames, IA, 50011, USA}
\affiliation{Department of Physics and Astronomy, Iowa State University, Ames, IA, 50011, USA}
\author{M. S. Scheurer}
\affiliation{Department of Physics, Harvard University, Cambridge, MA 02138, USA}
\affiliation{Institute for Theoretical Physics, University of Innsbruck, A-6020 Innsbruck, Austria}
\author{Y.M. Dai}
\affiliation{Center for Integrated Nanotechnologies, Los Alamos National Laboratory, Los Alamos, NM 87545, USA}
\affiliation{Center for Superconducting Physics and Materials, National Laboratory of Solid State Microstructures and Department of Physics, Nanjing University, Nanjing 210093, China}
\author{M.-C. Lee}
\affiliation{Center for Integrated Nanotechnologies, Los Alamos National Laboratory, Los Alamos, NM 87545, USA}
\author{P. Padmanabhan}
\affiliation{Center for Integrated Nanotechnologies, Los Alamos National Laboratory, Los Alamos, NM 87545, USA}
\author{L.T. Mix}
\affiliation{Center for Integrated Nanotechnologies, Los Alamos National Laboratory, Los Alamos, NM 87545, USA}
\author{S. W. Teitelbaum}
\affiliation{Stanford PULSE Institute, SLAC National Accelerator Laboratory, Menlo Park, California 94025, USA}
\affiliation{Stanford Institute for Materials and Energy Sciences, SLAC National Accelerator Laboratory, Menlo Park, California 94025, USA}
\author{M. Trigo}
\affiliation{Stanford PULSE Institute, SLAC National Accelerator Laboratory, Menlo Park, California 94025, USA}
\affiliation{Stanford Institute for Materials and Energy Sciences, SLAC National Accelerator Laboratory, Menlo Park, California 94025, USA}
\author{L.X. Zhao}
\affiliation{Institute of Physics, Chinese Academy of Sciences, Beijing 100190, China}
\author{G.F. Chen}
\affiliation{Institute of Physics, Chinese Academy of Sciences, Beijing 100190, China}
\author{B. Xu}
\affiliation{Institute of Physics, Chinese Academy of Sciences, Beijing 100190, China}
\author{R. Yang}
\affiliation{Institute of Physics, Chinese Academy of Sciences, Beijing 100190, China}
\author{B. Shen}
\affiliation{Department of Physics and Astronomy, University of California, Los Angeles, CA 90095, USA}
\author{C. Hu}
\affiliation{Department of Physics and Astronomy, University of California, Los Angeles, CA 90095, USA}
\author{C.-C. Lee}
\affiliation{Department of Physics, Tamkang University, Tamsui, New Taipei 251301, Taiwan}
\author{H. Lin}
\affiliation{Institute of Physics, Academia Sinica, Taipei 11529, Taiwan}
\author{T.A. Cochran}
\affiliation{Laboratory for Topological Quantum Matter and Advanced Spectroscopy (B7), Department of Physics, Princeton University, Princeton, New Jersey 08544, USA}
\author{S.A. Trugman}
\affiliation{Center for Integrated Nanotechnologies, Los Alamos National Laboratory, Los Alamos, NM 87545, USA}
\author{J.-X. Zhu}
\affiliation{Center for Integrated Nanotechnologies, Los Alamos National Laboratory, Los Alamos, NM 87545, USA}
\author{M.Z. Hasan}
\affiliation{Laboratory for Topological Quantum Matter and Advanced Spectroscopy (B7), Department of Physics, Princeton University, Princeton, New Jersey 08544, USA}
\affiliation{Materials Sciences Division, Lawrence Berkeley National Laboratory, Berkeley, California 94720, USA}
\author{N. Ni}
\affiliation{Department of Physics and Astronomy, University of California, Los Angeles, CA 90095, USA}
\author{X.G. Qiu}
\affiliation{Institute of Physics, Chinese Academy of Sciences, Beijing 100190, China}
\author{A.J. Taylor}
\affiliation{Center for Integrated Nanotechnologies, Los Alamos National Laboratory, Los Alamos, NM 87545, USA}
\author{D.A. Yarotski}
\affiliation{Center for Integrated Nanotechnologies, Los Alamos National Laboratory, Los Alamos, NM 87545, USA}
\author{R.P. Prasankumar}
\email{rpprasan@lanl.gov}
\affiliation{Center for Integrated Nanotechnologies, Los Alamos National Laboratory, Los Alamos, NM 87545, USA}

%\date{\today}% It is always \today, today,
             %  but any date may be explicitly specified

%\begin{abstract}
 % Symmetry plays a key role in both conventional and topological phases of matter, making the ability to optically drive symmetry changes a critical step in developing future technologies that rely on such control. Here, we use femtosecond optical pulses to transiently lower symmetry in the prototypical type-I Weyl semimetal TaAs. Using second harmonic generation spectroscopy, we observe an ultrafast reduction of magnetic point group symmetry without any structural change, indicating the electronic origin of the transition. We argue that this effect is brought on by photocurrent generation, which breaks both spatial and time-reversal symmetry through introducing an asymmetry in the non-equilibrium distribution of charge carriers following photoexcitation. Our results demonstrate that optically driven photocurrents explicitly break electronic symmetry in a generic fashion, opening up the possibility of driving phase transitions between symmetry-protected states on ultrafast time scales.
  %Our results underscore the role that photocurrents play in manipulating magnetic symmetries of electronic systems over ultrashort timescales to ultimately drive phase transitions between symmetry-protected states.
%\end{abstract}

%\keywords{Suggested keywords}%Use showkeys class option if keyword
                              %display desired
\maketitle

\section{Introduction} 
  \textbf{Symmetry plays a central role in conventional and topological phases of matter, making the ability to optically drive symmetry changes a critical step in developing future technologies that rely on such control. Topological materials, like the newly discovered topological semimetals, are particularly sensitive to a breaking or restoring of time-reversal and crystalline symmetries, which affect both bulk and surface electronic states. While previous studies have focused on controlling symmetry via coupling to the crystal lattice~\cite{Sie_Symmetry_2019,Collins_DSM_FET_2018,Mutcheaav9771_ZrTe5,PhysRevX.10.021013,Luo2021}, we demonstrate here an all-electronic mechanism based on photocurrent generation. Using second-harmonic generation spectroscopy as a sensitive probe of symmetry changes ~\cite{Torchinsky2017,Zhao_SHGReview_2018}, we observe an ultrafast breaking of time-reversal and spatial symmetries following femtosecond optical excitation in the prototypical type-I Weyl semimetal TaAs. Our results show that optically driven photocurrents can be tailored to explicitly break electronic symmetry in a generic fashion, opening up the possibility of driving phase transitions between symmetry-protected states on ultrafast time scales.}
  
  Symmetry breaking has long defined the dominant paradigm for describing phase transitions in condensed matter systems. More recently, the discovery of novel topological phases, characterized by topological invariants as opposed to a local order parameter arising from spontaneously broken symmetry, provides an alternative framework for classifying states of matter \cite{WenFQH,SubirReview}. Nevertheless, symmetry continues to play a central role in the physics of topological materials, as it underlies topological protection in topological insulators and superconductors \cite{Hasan_RMP_2010}, crystalline topological phases \cite{TCIReview}, and the recently discovered topological semimetals \cite{Gao_ARMR_2019, Armitage2018, Hasan2017, Yan2017}. In Dirac semimetals, symmetry protects the four-fold degeneracy of the Dirac point \cite{Young_DSM_Sym_2012}, while for Weyl semimetals (WSMs), the breaking of time-reversal or inversion symmetry allows for the crossing of two linearly dispersing, non-degenerate bands, giving rise to Weyl points \cite{Wan2011, Xu2011, Xu2015, Lv2015, Yang2015}. These points act as monopoles of Berry curvature in momentum ($k$) space, and their presence leads to several unique experimental manifestations \cite{Xu2015, Lv2015, Yang2015, Liu2015, Jia2016, Parameswaran2014, Huang2015, Zhang2016} that make these materials appealing for future technological applications \cite{Tokura2017}.

\begin{figure*}
  \centering
      \includegraphics[width=\textwidth]{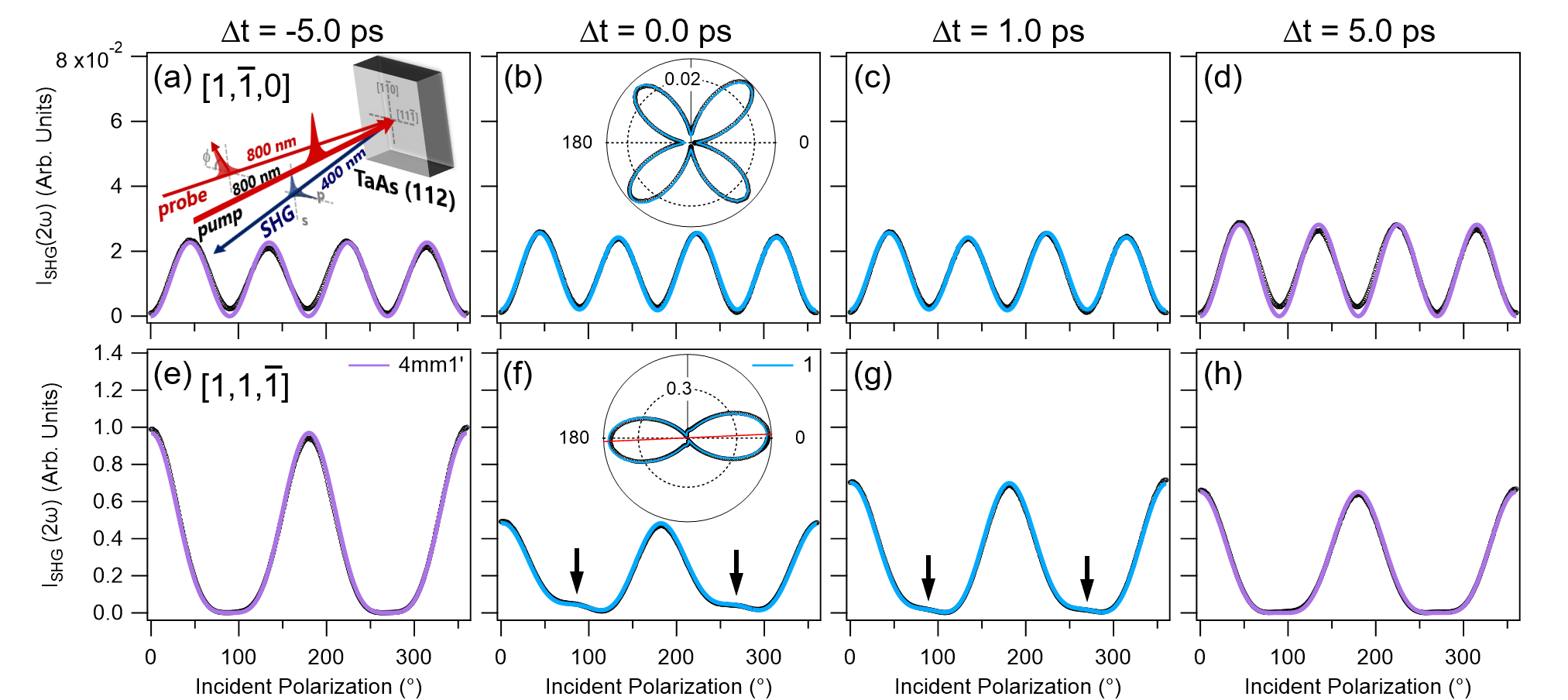}
      \caption{Snapshots of the SHG pattern ($2\hbar\omega\sim3.1$ eV) measured along the (a-d)  [1,$\bar{1}$,0] and (e-h) [1,1,$\Bar{1}$] axes for various pump delays: (a,e) $\Delta t= -$5.0 ps, (b,f) 0 ps, (c,g) 1.0 ps and (d,h) 5.0 ps. The inset in (a) shows a schematic of the experimental geometry, while insets in (b) and (f) show polar plots of the SHG pattern immediately following a linearly polarized pump excitation ($\hbar \omega$= 1.55 eV; fluence = 4.34 mJ/cm$^{2}$) aligned nearly along the [1,1,$\bar{1}$]) axis. Fits of the pattern assuming the magnetic point symmetries of $4mm1'$ ((a,e);(d,h)) and $1$ ((b,f);(c,g)) are shown as solid purple and blue traces, respectively. Arrows in (f) and (g) denote the presence of transient, asymmetric lobes in the photoexcited state, while a small $\sim 2.5^\circ$ rotation of the SHG pattern along [1,1,$\bar{1}$] is evident from the inset in (f).}
      \label{fig:TR_RASHG}
  \end{figure*}
  
  Conventional probes of symmetry rely on diffractive techniques, like x-ray, neutron, and electron scattering, to determine the respective lattice, magnetic, and charge ordering in a crystal. Nonlinear optics is also an effective probe of symmetry, as the nonlinear response is described by a third (or higher) rank tensor \cite{Torchinsky2017,Zhao_SHGReview_2018}, allowing for phases hidden to linear probes (e.g., in correlated electron systems) to be revealed \cite{Zhao_SHG_2016,Harter_SHG_2017,VanAken2007,Jin2020}. In the transition metal monopnictide (TMMP) family of WSMs, the lack of inversion symmetry resulting from a polar c-axis leads to an especially strong nonlinear optical response, with significant contributions from the generation of helicity-dependent injection \cite{Ma2017,Sirica_THz_2019,Gao_ChiralTHz_2020} and helicity-independent shift \cite{Wu_SHG_2016,Patankar_SHG_2018,Osterhoudt2019,Ma2019,Sirica_THz_2019} photocurrents. Shift currents, resulting from a coherent shift of the electron cloud in real space following photoexcitation \cite{Sipe2000}, are particularly important, as they play a dominant role in both the giant, anisotropic second harmonic response \cite{Wu_SHG_2016,Patankar_SHG_2018,Li2018} as well as the bulk photovoltaic effect \cite{Osterhoudt2019} seen in WSMs, and may be traced to a difference in Berry connection between the bands participating in the optical transition \cite{Morimoto2016,Sotome_ShiftCurrent_2019}. The most common nonlinear optical probe, second harmonic generation (SHG) spectroscopy, is thus sensitive to the asymmetric carrier distribution that accompanies photocurrent generation, making it a powerful tool for measuring the effect of transient photocurrents on material symmetry.
  
  In this Letter, we show that femtosecond (fs) optical excitation transiently lowers the magnetic point symmetry $4mm1'$, with $1'$ indicating time-reversal symmetry, of the type-I WSM TaAs. Time-resolved SHG (TR-SHG) spectroscopy reveals this symmetry change occurs on a picosecond (ps) timescale, with no accompanying structural transition, indicating it to be purely electronic in origin. The strong nonlinear optical response exhibited by the TMMP WSMs \cite{Morimoto2016,Wu_SHG_2016,Patankar_SHG_2018,Ma2017,Osterhoudt2019,Parker_NLO_Diagram_2019,Sirica_THz_2019,Gao_ChiralTHz_2020,Weber_MPTP_2017,Weber2021} allows us to attribute this reduction in symmetry to changes in the spatial distribution of the electronic polarization that follow from photocurrent generation, supported by our previous terahertz (THz) emission experiments \cite{Sirica_THz_2019}. The degree of symmetry breaking is governed by the current direction, which we manipulate via the pump polarization. Our results demonstrate that optically driven photocurrents generically break electronic symmetries and can be used to achieve dynamic control of material properties on ultrafast time scales. This control mechanism will have wide ranging applications, particularly for topological semimetals, where symmetry is intimately tied to topology, opening up an original %new 
  avenue of study rooted in current-induced symmetry breaking \cite{Khurgin_1995, Ruzicka_PRL_2012}. %\cite{Kazuaki2020}.   
  %Our results demonstrate optically driven photocurrents that break electronic symmetries in this fashion can be used to achieve dynamic control of material properties on ultrafast time scales. 
 
\section{Results and Discussion}
  Prior to pump excitation, Figs.~\ref{fig:TR_RASHG}(a) and (e) show SHG patterns collected along the two in-plane [1,$\bar{1}$,0] and [1,1,$\bar{1}$] axes of the (112) face that are well described by a nonlinear susceptibility tensor, $\chi^{(2)}_{ijk}(2\omega)$, obeying the $4mm1'$ point group symmetry of TaAs (Section 10 of the supplementary information~\cite{Supp}). Here, the emitted second harmonic is dominated by an electric dipole response that is attributed to the polar $c$-axis. This is reflected by a large ratio of $\chi^{(2)}_{zzz}/\chi^{(2)}_{xxz}=7.4$ at $\hbar\omega = 1.55$ eV (for $z$ parallel to the crystallographic $c$-axis), as determined from our fits and in agreement with Refs.~\onlinecite{Wu_SHG_2016} and \onlinecite{Patankar_SHG_2018}. 

\begin{figure*}[t!]
  \centering
      \includegraphics[width=\textwidth]{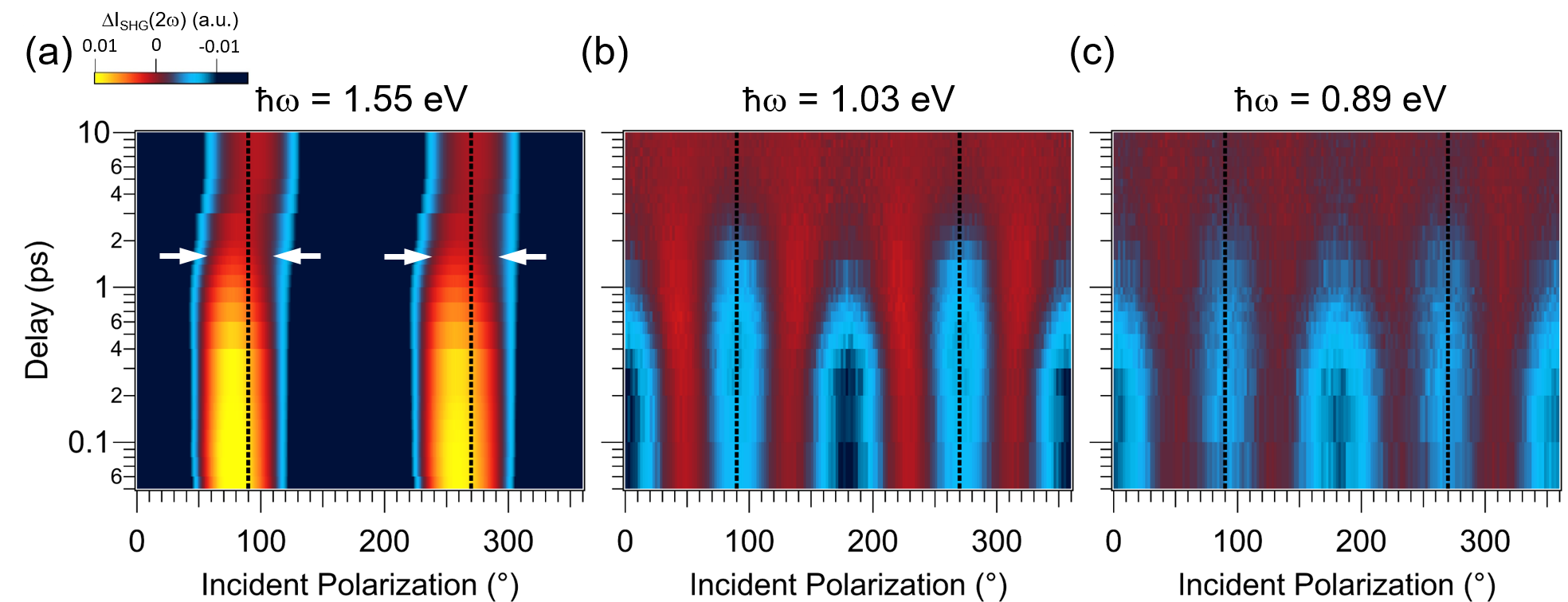}
      \caption{Photoinduced polarization- and time-dependent changes in SHG intensity, $\Delta I_{\text{SHG}}(2\omega)$, measured across the entire [1,1,$\Bar{1}$] pattern after 1.55 eV photoexcitation (fluence = 4.34 mJ/cm$^{2}$) using probe energies of (a) $\hbar\omega$ = 1.55 eV, (b) $\hbar\omega$ = 1.03 eV, and (c) $\hbar\omega$ = 0.89 eV (Fig. S1 \cite{Supp}). Arrows in (a) denote the recovery of $4mm1'$ symmetry following the decay of the emergent, asymmetric lobes at $\sim90^{\circ}$ and $\sim270^{\circ}$ as well as a rotation of the SHG pattern back to equilibrium. Compared to panel (a), optical pumping at 1.55 eV does not change the symmetry of the patterns in panels (b-c), but suppresses them in a nearly isotropic manner over a $\Delta t< 2.0$ ps timescale (Fig. S2 \cite{Supp}).}
      \label{fig:PI_RASHG}
  \end{figure*}
  
  Following 1.55 eV photoexcitation, Figs.~\ref{fig:TR_RASHG}(f-h) show pronounced changes in the SHG pattern along [1,1,$\Bar{1}$] that can be traced to a transient change in symmetry within the material when the pump polarization is nearly aligned along the [1,1,$\Bar{1}$] axis. With the arrival of the pump pulse in Fig.~\ref{fig:TR_RASHG}(f), the emitted SHG along [1,1,$\Bar{1}$] is reduced by half and the resultant pattern exhibits a $2.5^\circ$ rotation with respect to equilibrium (inset of Fig.~\ref{fig:TR_RASHG}(f)). In addition, small lobes absent from the static pattern appear at $\sim 90^\circ$ and $\sim 270^\circ$ (Figs.~\ref{fig:TR_RASHG}(f-g)), whose asymmetry suggests a reduction of symmetry in the photoexcited state. In contrast, SHG patterns along [1,$\bar{1}$,0] grow in amplitude, with no additional rotation or spectral features appearing under pump excitation (Figs.~\ref{fig:TR_RASHG}(b-d)). From Fig.~\ref{fig:TR_RASHG}(g), both the rotation and asymmetric lobes in the [1,1,$\Bar{1}$] TR-SHG pattern follow similar ultrafast dynamics, lasting $\tau_{PI}\sim 1.1$ ps before symmetry is restored and the intensity of the main lobes at $0^\circ$ and $180^\circ$ begins to recover back to its equilibrium value (Fig.~\ref{fig:TR_RASHG}(h)). 

  Coupling of the dynamics for both the rotation and lobe asymmetry in the TR-SHG spectra is further illustrated in Fig.~\ref{fig:PI_RASHG}, showing photoinduced polarization- and time-dependent changes over the entire [1,1,$\Bar{1}$] pattern taken for the three probe energies used in our experiments. Fig.~\ref{fig:PI_RASHG}(a) reveals that both spectral features exhibit an equivalent time dependence, suggesting they originate from the same photoinduced symmetry-breaking transition, while Figs.~\ref{fig:PI_RASHG}(b-c) reveal an absence of symmetry breaking under non-resonant probe conditions (discussed further below). Together with separate time-resolved X-ray diffraction experiments (Fig. S3~\cite{Supp,MCLee2020}), which show no structural dynamics over ultrafast timescales, but only on significantly longer timescales due to laser heating, this suggests an electronic origin of the symmetry breaking transition. Additionally, Fig.~\ref{fig:PI_RASHG}, along with symmetry considerations (Section 10~\cite{Supp}), excludes the possibility that a dominant surface contribution, arising from a screened bulk response due to a high density of photoexcited carriers ($~10^{19}$ - $~10^{20}$ cm$^{-3}$), is responsible for the reduced symmetry state, as such an effect would be evident at all probe energies. Hence, the mere generation of a photoexcited charge density is insufficient for lifting $4mm1'$ symmetry, and it is only when resonantly probing the transiently excited state that symmetry breaking in the SHG pattern is observed. 
  
\begin{figure}[t!]
      \centering
      \includegraphics[width=\linewidth]{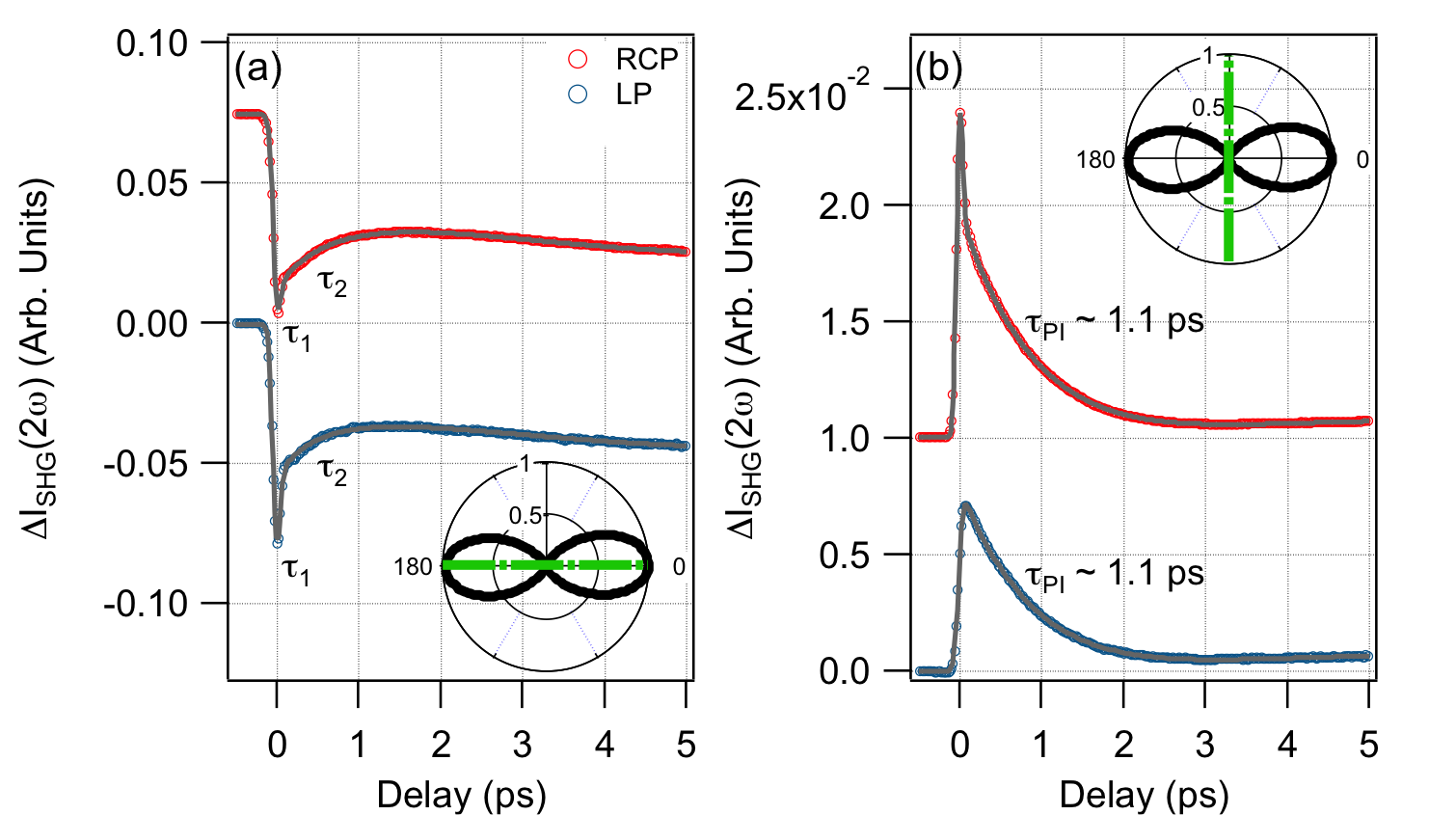}
      \caption{Time-dependent traces of $\Delta I_{\text{SHG}}(2\omega)$, measured for input probe polarizations along the green dashed lines of the [1,1,$\Bar{1}$] SHG pattern shown as insets following 1.55 eV pump excitation. Here, (a) the suppression of the main lobe and (b) emergence of the photoinduced, asymmetric lobe are captured with high temporal resolution ($<100$ fs). Fits of the dynamics following (blue) linearly and (red) circularly polarized pump excitation at a fluence of 4.34 mJ/cm$^{2}$ are superimposed onto the data, where traces generated under circularly polarized excitation have been offset for clarity. Traces in (a) show a pulsewidth-limited ultrafast component ($\tau_{1}\sim$ 80 fs) which is weakly dependent on pump polarization, while a similar ultrafast component develops only under helicity-dependent photoexcitation in (b) and arises from a sub-100 fs dichroic response (Fig. S4~\cite{Supp}).} 
      \label{fig:TR_SHG_Trace}
  \end{figure}

  By reducing symmetry, the constraints imposed in equilibrium are lifted, necessitating that we consider a lower symmetry sub-group of $4mm1'$ to describe the time-dependent nonlinear susceptibility elements $\chi^{(2)}_{ijk}(2\omega; \Delta t)$, where $\omega$ and $\Delta t$ denote frequency and time delay after pump excitation, respectively. To quantitatively extract information about the behavior of $\chi^{(2)}_{ijk}(2\omega; \Delta t)$, we simultaneously fit the SHG patterns collected along [1,$\Bar{1}$,0] and [1,1,$\Bar{1}$] as a function of pump delay and incident polarization angle, $\phi$ (Fig.~\ref{fig:TR_RASHG}). As compared to equilibrium, the rotation of the pattern along [1,1,$\Bar{1}$] and the emergence of asymmetric lobes at $90^{\circ}$ and $270^{\circ}$ following photoexcitation cannot be accounted for under $4mm1'$ symmetry, as this requires the lobes to be both symmetric and pinned along the $x$ and $y$ axes. Rather, by considering the different subgroups of $4mm1'$, we find an optimal fit that captures the aforementioned features of the photoexcited state only in the absence of time-reversal and diagonal mirror, $m_{x,x,z}$, symmetry, described by the magnetic point group $1$ (no point symmetries). Using the expression $I^\alpha_{\text{SHG, 1}}(\phi) = \sum_{n=0}^4 \mathcal{C}_n^{\alpha} \sin^{n}(\phi) \cos^{4-n}(\phi) $ ($\alpha = [1\bar{1}0], [11\bar{1}]$) for $1$ symmetry over a $\Delta t < 2.0$ ps timescale allows us to associate different fit coefficients, $\mathcal{C}_n^{[11\bar{1}]}$, to specific features in the pattern (Section 10~\cite{Supp}). The dynamics of these features are captured by the TR-SHG traces in Fig.~\ref{fig:TR_SHG_Trace} ($\Delta I_{\text{SHG}}(2\omega)$), plotted as a function of pump delay for select combinations of input and output probe polarizations. Fig.~\ref{fig:TR_SHG_Trace}(a) captures the time dependence of the largest fit coefficient $\mathcal{C}_0^{[11\bar{1}]}$, depicting the suppression and subsequent recovery of the dominant lobe in the SHG pattern on a timescale defined by $\tau_{1}$ and $\tau_{2}$, while Fig.~\ref{fig:TR_SHG_Trace}(b) shows the time dependence of $\mathcal{C}_4^{[11\bar{1}]}$, which illustrates the dynamics of the emergent, asymmetric lobe arising from photoexcitation.
  
  Despite both $\mathcal{C}_{0}^{[11\bar{1}]}$ and $\mathcal{C}_{4}^{[11\bar{1}]}$ being allowed under $4mm1'$ symmetry (though $\mathcal{C}_4^{[11\bar{1}]}$ is small for the static pattern at $\hbar\omega =1.55$~eV), a reduction to $1$ symmetry is captured by the pair of odd fit parameters, $\mathcal{C}^{[11\bar{1}]}_{1}$ and $\mathcal{C}^{[11\bar{1}]}_{3}$, that quantitatively measure the degree of symmetry breaking imposed by the pump following photoexcitation (Section 10~\cite{Supp}). These parameters result from breaking $m_{x,x,z}$ and enable fitting of the overall rotation in the pattern. However, as long as time-reversal symmetry remains, relations between the different $\mathcal{C}^{[11\bar{1}]}_{n}$ prevent these odd coefficients from capturing the observed asymmetry seen in our data. Rather, an accurate description requires a breaking of time-reversal symmetry to remove the constraints imposed on $\mathcal{C}^{[11\bar{1}]}_{n}$, revealing that both $m_{x,x,z}$ and time-reversal symmetry must be lifted in order to fully describe our experimental results. This reduction of symmetry from $4mm1'$ to $1$ also allows for the emergence of additional fit parameters in the [1,$\Bar{1}$,0] pattern, but these remain small and time-independent, consistent with the weaker ($\sim100\times$) nonlinear response observed along this axis.
 
  \begin{figure}[t]
      \centering
      \includegraphics[width=\linewidth]{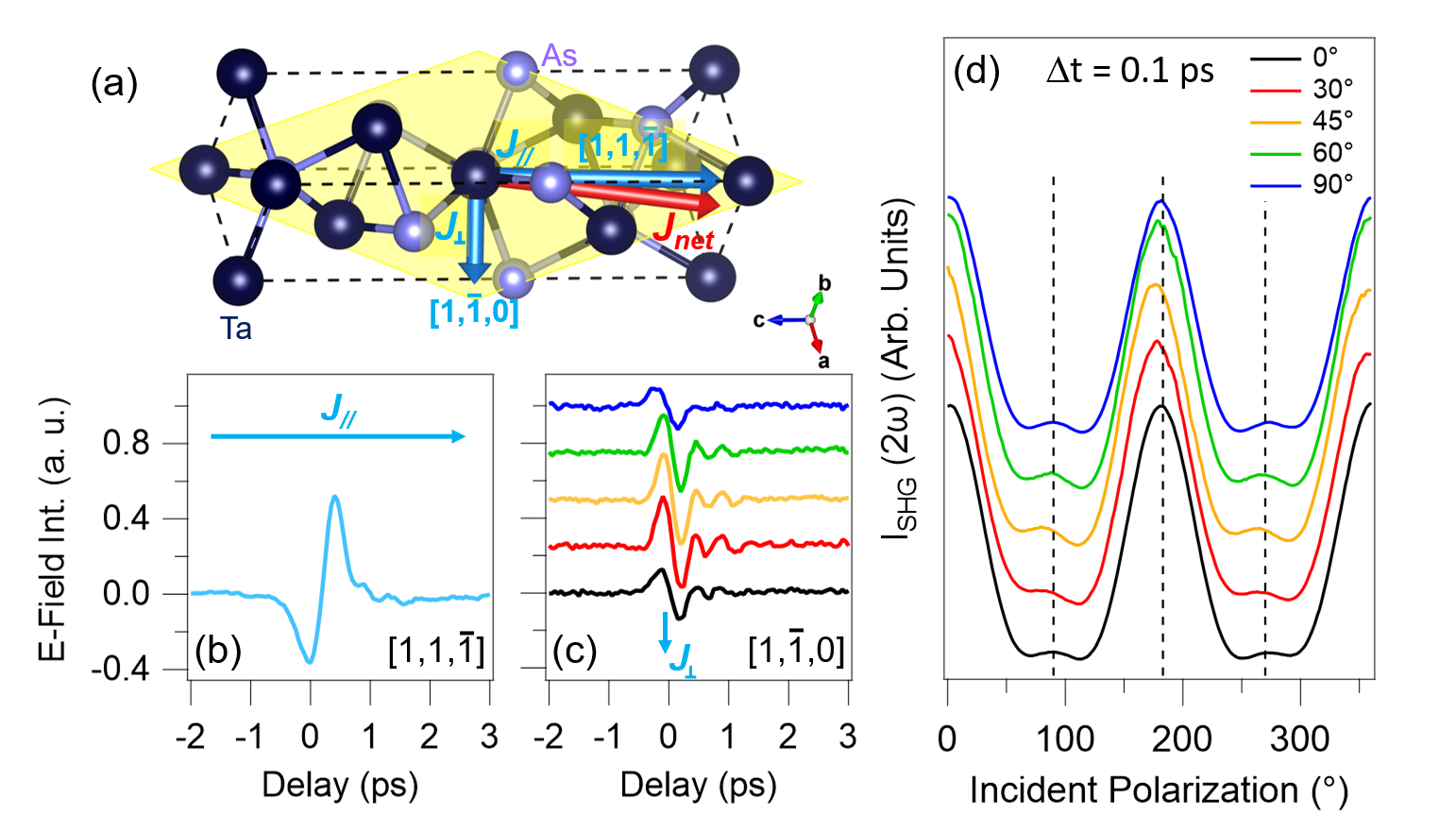}
      \caption{(a) Schematic diagram of the net photocurrent contained within the (112) plane (yellow) of a TaAs unit cell. Emitted THz waveforms resulting from transient photocurrents generated along the (b) [1,1,$\bar{1}$] and (c) [1,$\bar{1}$,0] axes (offset for clarity) under linearly polarized pump excitation. (d)~Changes in the transient SHG pattern ($\Delta t\sim 0.1$ ps) measured as a function of pump polarization relative to the [1,1,$\Bar{1}$] axis (offset for clarity). The presence of an enhanced ($\sim10\% -20\%$), polarization-dependent shift current along [1,$\bar{1}$,0] leads to a clear reduction in symmetry within the pattern (Section 9 \cite{Supp}).}
      \label{fig:Photocurrent}
  \end{figure}
  
  We emphasize that in contrast with the spontaneous symmetry breaking seen in conventional photoinduced phase transitions ~\cite{Zong_TD_2019, Zong_DSD_2019}, the symmetry breaking observed here is explicit and originates from the photoexcitation process itself. In this regard, the symmetry resolution gained from nonlinear optical probes like SHG provides a completely generic and robust framework for characterizing light-induced changes in the non-equilibrium state, as no underlying assumptions or reliance on theoretical models are required~\cite{Torchinsky2017,Zhao_SHGReview_2018}. While previous studies have reported explicit symmetry breaking from polarization-dependent photoexcitation in Bi and Sb that results from the coupling of phonons to a transiently excited charge density in $k$-space~\cite{Li_PI_Bi_2013, Murray_SymmetryBreaking_2015, O'Mahony_SymmetryBreaking_2019}, we propose an alternative mechanism, which is entirely electronic in origin. Accounting for the fact that optical excitation creates a highly non-equilibrium state, we have carefully considered a number of possibilities, including a spatially inhomogeneous pump volume, anisotropic changes in optical constants, and carrier thermalization following photoexcitation, before concluding that the lowering of $4mm1'$ symmetry to $1$ in TaAs most likely arises from photocurrent generation (Section 11 \cite{Supp}).
   
  By having a well defined, but generic, direction relative to some high symmetry axis of the crystal, a photocurrent, $\Vec{J}$, breaks both spatial and time-reversal symmetry through introducing an asymmetry in the non-equilibrium distribution of charge carriers along its direction that must necessarily be odd under time reversal. On the (112) face of TaAs, photocurrents originating from asymmetry in the real (shift) or $k$-space (injection) carrier density are allowed to flow along the [1,1,$\Bar{1}$] and [1,$\Bar{1}$,0] axes \cite{Sirica_THz_2019,Gao_ChiralTHz_2020}. Here, a reduction in symmetry from $4mm1'$ to $1$ occurs when a net current is directed away from either of these two high symmetry axes, breaking diagonal mirror symmetry (Fig.~\ref{fig:Photocurrent}(a)) (Section 10~\cite{Supp}). While a polarization-independent photocurrent is always present along [1,1,$\bar{1}$] (Fig.~\ref{fig:Photocurrent}(b)) \cite{Sirica_THz_2019}, the polarization dependence of the [1,$\Bar{1}$,0] photocurrent allows us to break symmetry in a controllable manner by exploiting symmetry constraints placed on the shift current following linearly polarized excitation (Fig.~\ref{fig:Photocurrent}(c)) (Section 9~\cite{Supp}). As shown in Fig.~\ref{fig:Photocurrent}(d), varying the linear polarization of the excitation pulse relative to the [1,1,$\Bar{1}$] axis causes the emergent photoinduced lobes in the transient SHG pattern ($\Delta t \sim  0.1$ ps) to develop a clear polarization-dependent asymmetry, while the pattern itself exhibits a rotation with respect to $0^{\circ}$ and $180^{\circ}$. This is fully consistent with a reduction of symmetry brought on by an enhanced shift current response; as the pump polarization is detuned from the [1,1,$\Bar{1}$] axis, shift currents along [1,$\Bar{1}$,0] become symmetry allowed, with the largest contribution coming from an equal projection of the pump polarization along the two orthogonal axes of the (112) face (i.e. $45^{\circ}$) (Section 9~\cite{Supp}). This is reproduced by our data in Fig.~\ref{fig:Photocurrent}(d), and illustrates our ability to exploit symmetry constraints on the photocurrent to tune the degree of symmetry breaking in this material. Similarly, after circularly polarized excitation, symmetry dictates that injection photocurrents can only flow along [1,$\Bar{1}$,0], leading to an ultrafast reduction of $4mm1'$ to $1$, as confirmed in Fig.~\ref{fig:TR_SHG_Trace}(b) and Fig.~S4 \cite{Supp}.   
  
  \emph{Ab initio} calculations for the optical conductivity in TaAs reveal an As-$p$ to Ta-$d$ transition to be the major contribution to the in-plane conductivities, $\sigma_{xx}$ and $\sigma_{yy}$, for our 1.55~eV excitation energy, while a Ta-$d$ to Ta-$d$ transition dominates the out-of-plane response, $\sigma_{zz}$ (Section 12~\cite{Supp}). The fact that no symmetry changes are observed under non-degenerate pump excitation for either $\hbar\omega_{pump}>\hbar\omega_{probe}$ (Fig.~\ref{fig:PI_RASHG}(b-c)) or $\hbar\omega_{pump}<\hbar\omega_{probe}$ (Fig. S5~\cite{Supp}) can thus be attributed to resonantly probing this initial photoexcited population of carriers. In other words, the transient asymmetry generated under 1.55~eV photoexcitation is lost over an ultrafast timescale due to momentum scattering, consistent with the bandwidth of the emitted THz pulses \cite{Sirica_THz_2019}, and is therefore absent as photoexcited carriers relax to the lower lying, non-degenerate energy states probed at 0.89 eV - 1.03 eV (Fig.~\ref{fig:PI_RASHG}(b-c)). Further support is provided by band-resolved imaging of the photocurrent response in the topological insulator Bi$_2$Se$_3$, which reveals the transient asymmetry due to optically excited photocurrents is lost within $\sim$165 fs \cite{Soifer_BRPC_2019}. Hence, by using degenerate pump and probe energies, we ensured that only those states responsible for generating the photocurrent following pump excitation are probed; we note that any contribution arising from sum frequency generation can be discounted, since symmetry breaking in the TR-SHG pattern exists on timescales much longer than the temporal overlap of the pump and probe pulses. 
  
  Finally, our particular choice of photon energy was based on the giant anisotropic nonlinear response of the static SHG pattern along [1,1,$\Bar{1}$] at 1.55~eV \cite{Wu_SHG_2016,Patankar_SHG_2018}. While degenerate TR-SHG experiments performed at 1.02 eV show some dependence on pump polarization, similar to Fig.~\ref{fig:Photocurrent}, the photoinduced change across the entire SHG pattern is dominated by a suppression of the SHG response, with the pattern itself retaining $4mm1'$ symmetry within our experimental resolution (Fig. S6~\cite{Supp}). This is unsurprising, since the lack of spectral features at $90^{\circ}$ and $270^{\circ}$ in the static SHG pattern at 1.55 eV (Fig.~\ref{fig:TR_RASHG}(e)) as compared to lower photon energies (Fig. S1 (b-c)), makes this probe photon energy optimal for observing photocurrent-induced SHG along the orthogonal [1,$\Bar{1}$,0] axis, as manifested by the emergent photoinduced lobes in Fig.~\ref{fig:TR_RASHG}(f-g). 
  
  In conclusion, by performing TR-SHG spectroscopy on the (112) surface of the WSM TaAs, we reveal a transient breaking of all magnetic point group symmetries following optical excitation, reducing the symmetry from $4mm1'$ to $1$. Both the prompt recovery of equilibrium symmetry, as well as the absence of an ultrafast structural transition following optical excitation, suggest that light-induced symmetry breaking in TaAs originates from transient photocurrent generation. Specifically, the presence of a polar $c$-axis in the TMMP WSMs leads to a dominant helicity-independent, shift current whose geometric interpretation is rooted in an asymmetry in the electronic polarization introduced by optical excitation \cite{Patankar_SHG_2018}. In this regard, our TR-SHG study reflects time-dependent changes to the polarization distribution that fail to respect both spatial and time-reversal symmetries, and whose relaxation is governed by a polarization-independent recovery, $\tau_{PI}$, describing the return of the electronic polarization back to equilibrium. %We further demonstrate that the degree of symmetry lowering can be controlled via changing the pump polarization.
  
  The effect we report in this study originates from the fact that symmetry imposes general constraints on material properties, and thus can have important consequences for topological materials, where topology is closely related to symmetry. Since symmetry constrains the total number of Weyl nodes in TaAs, a reduction in symmetry brought on by photocurrent generation is expected to shift these nodes in both energy and momentum. This will alter the Fermi arc surface states, suggesting future time-and-angle-resolved photoemission spectroscopy experiments to directly measure the impact of transient photocurrents on the electronic band topology. More generally, our findings can be applied to any material system where either an optically generated or externally applied current breaks electronic symmetries. While generic, these results have important implications for topological semimetals \cite{Kazuaki2020}, as the ability to alter symmetry on ultrafast timescales in these materials can lead to the potential realization of topological field effect transistors. 

\section{Methods}
\subsection{Crystal Growth}
  TaAs single crystals were grown from polycrystalline samples by chemical vapor transport using iodine (2 mg/cm$^3$) as the transporting agent. Large polyhedral crystals with dimensions up to 1.5 mm were obtained in a temperature field of $\Delta$T = 1150$-$1000$^{\circ}$C following 3 weeks at growth temperature in an evacuated quartz ampoule. The as-grown three-dimensional (3D) crystals exhibit multiple surface facets, with the (112) face being identified by X-ray diffraction measurements.
    
\subsection{Time-resolved SHG}
  TR-SHG experiments were performed on the (112) surface of different as-grown TaAs single crystals, sourced from entirely different batches, using an amplified Ti:Sapphire laser system operating at a 250 kHz repetition rate. SHG generated at near normal incidence ($6^{\circ}$) from a linearly polarized optical probe tuned over a 0.89-1.55 eV (800 nm - 1400 nm) energy range was measured as a function of incident light polarization. A Glan-Taylor polarizer was used to select the emitted second harmonic polarized along the in-plane [1,$\Bar{1}$,0] and [1,1,$\Bar{1}$] high symmetry axes of the (112) face, as described in Ref. \onlinecite{Sheu2014}. Here, initial crystal alignment was performed by Laue diffraction, allowing us to hand scribe the direction of these two high symmetry axes (parallel to crystal face edge) on the sample mount ($\pm~5^{\circ}$). Static and time-resolved SHG measurements were carried out in air or an optical cryostat, with the [1,1,$\Bar{1}$] axis oriented parallel to the optical table. Initial pump and probe polarizations were aligned as closely to the [1,$\Bar{1}$,0] and [1,1,$\Bar{1}$] axes as possible, and verified through comparing the static SHG patterns measured here with those in literature \cite{Wu_SHG_2016,Patankar_SHG_2018}. 
  
  For excitation fluences ranging from 0.48 mJ/cm$^{2}$ - 6.03 mJ/cm$^{2}$ (Fig. S7 \cite{Supp})), a normally incident, $\sim 80$ fs optical pump pulse centered at 1.55 eV and either circularly or linearly polarized with respect to the [1,1,$\Bar{1}$] axis creates a photocarrier density of $~10^{19}$ - $~10^{20}$ cm$^{-3}$ within the 22 nm penetration depth \cite{Xu2016}. In this pump fluence range, temperature (5-300 K) (Fig. S8 \cite{Supp}) and pump polarization-dependent TR-SHG traces were measured for select combinations of input and output probe polarizations, while photoinduced changes across the entire SHG pattern were likewise collected at specific pump delays. In these experiments, a pump beam diameter $1.4\times$ larger than the probe ensures that an initially homogeneous distribution of photoexcited carriers is measured. Finally, the generation of a photocurrent under pump excitation was ensured through working in the appropriate fluence range as well as observing THz-field induced second harmonic generation (TFISH) in the [1,$\Bar{1}$,0] pattern, confirming the presence of a photocurrent under optical excitation \cite{Sirica_THz_2019}.
 
\subsection{Ab initio Calculations}
  First-principles calculations were performed using the OpenMX code, where norm-conserving pseudopotentials, optimized pseudoatomic basis functions, and the generalized gradient approximation were adopted \cite{Abinitio_PhysRevB.67.155108_01,abinitio_PhysRevB.47.6728_02,GGA_PhysRevLett.77.3865}. Spin-orbit coupling was incorporated in these calculations through the use of j-dependent pseudopotentials \cite{LS_PhysRevB.64.073106}. For each Ta atom, three, two, two, and one optimized radial functions were allocated for the s, p, d, and f orbitals with a cutoff radius of 7 bohrs, respectively, denoted as Ta $7.0-$s3p2d2f1. For each As atom, As $9.0-$s3p3d3f2 was adopted. A cutoff energy of 1000 Ry was used for the numerical integration and for the solution of the Poisson equation. A $17\times17\times17$ $k$-point sampling in the first Brillouin zone was used, with experimental lattice parameters being adopted in these calculations. Such choice of parameters is consistent with those described elsewhere \cite{TaAs_Lee_PhysRevB.92.235104}. The density-of-states (DOS) was calculated using $80\times80\times80$ $k$-points, with a Gaussian broadening of $0.05$ eV, for a primitive unit cell containing four atoms. The optical conductivity was then calculated via the Kubo-Greenwood formula using pseudoatomic basis functions \cite{Fe_AHE_PhysRevB.98.115115}. Similarly, an $80\times80\times80$ $k$-point sampling, and a broadening parameter of $\eta = 0.05$ eV were adopted. For both of the ground-state and conductivity calculations, an electronic temperature of 300 K was used.
  
\section{Author Contributions}
  TaAs single crystals were grown and characterized by L.X.Z., G.F.C., B.X., R.Y., B.S., C. H., N.N., and X.G.Q., with additional sample characterization and physical insights provided by T.A.C. and M.Z.H. N.S. and Y.M.D. performed the TR-SHG experiments with help from M.-C.L., P.P., and L.T.M. N.S., S.W.T., M.T., and R.P.P. performed the TR-XRD experiments with help from LCLS staff. The data was analyzed by N.S., P.P.O., and M.S.S. with a detailed symmetry analysis performed by P.P.O. and M.S.S. Ab initio calculations were carried out by C.-C.L. and H.L. with additional insight provided by J.-X.Z. The manuscript was written by N.S., P.P.O., M.S.S., and R.P.P. with significant contributions from S.A.T., A.J.T. and D.A.Y. 
  
\section{Competing Interests}
  The authors declare no competing interests.

\section{Data Availability}
  Relevant material, data, and associated protocols, including code and scripts, are curated and archived at the Materials Data Facility (https://doi.org/10.18126/lram-eh2d), and made available to the public.

\section{Acknowledgements} 
 This work was performed at the Center for Integrated Nanotechnologies at Los Alamos National Laboratory (LANL), a U.S. Department of Energy, Office of Basic Energy Sciences user facility, under user proposals $\#$2017BC0064 and $\#$2019AU0167. Use of the Linac Coherent Light Source (LCLS), SLAC National Accelerator Laboratory, is supported by the U.S. Department of Energy, Office of Science, Office of Basic Energy Sciences under Contract No. DE-AC02-76SF00515. N.S. and R.P.P. gratefully acknowledge the support of the U.S. Department of Energy through the LANL LDRD Program. P.P.O., J.-X.Z., and D.A.Y. are supported by the Center for Advancement of Topological Semimetals, an Energy Frontier Research Center funded by the U.S. Department of Energy Office of Science, Office of Basic Energy Sciences, through the Ames Laboratory under its Contract No. DE-AC02-07CH11358.  T.A.C. and M.Z.H. acknowledge support from the U.S. Department of Energy under grant DE-FG-02-05ER46200. Work at UCLA was supported by the US DOE, Office of Science, Office of Basic Energy Sciences under award no. DE-SC0021117 for single crystal growth and characterization. C. H. thanks the support of the Julian Schwinger Fellowship at UCLA. M.S.S. acknowledges support from the National Science Foundation under Grant No. DMR-1664842. C.-C.L. acknowledges the Ministry of Science and Technology of Taiwan for financial support under contract No. MOST 108-2112-M-032-010-MY2. T.A.C. was supported by the National Science Foundation Graduate Research Fellowship Program under Grant No. DGE-1656466. M.T., and S.W.T. were supported by the U.S. Department of Energy, Office of Science, Office of Basic Energy Sciences through the Division of Materials Sciences and Engineering under Contract No. DE-AC02-76SF00515. We thank Yu-Miin Sheu for the helpful discussion.   
 
%\bibliography{Sirica_Main_v1.bib}% Produces the bibliography via BibTeX.

\onecolumngrid
\newpage

% Mathias' modifications
\renewcommand{\thefigure}{S\arabic{figure}} % to get rid of space between S and number
%\usepackage{rotating}

% Added by Peter
\renewcommand\theequation{S\arabic{equation}}
\renewcommand\thetable{S\arabic{table}}
\renewcommand\thesection{\arabic{section}}
\newcommand{\equref}[1]{Eq.~(S\ref{#1})}
\newcommand{\bfa}{{\boldsymbol{a}}}
\newcommand{\bfb}{{\boldsymbol{b}}}
\newcommand{\bfe}{{\boldsymbol{e}}}
\newcommand{\bfk}{{\boldsymbol{k}}}
\newcommand{\bfv}{{\boldsymbol{v}}}
\newcommand{\bfx}{{\boldsymbol{x}}}
\newcommand{\bfy}{{\boldsymbol{y}}}
\newcommand{\bfz}{{\boldsymbol{z}}}
\newcommand{\bfee}{{\boldsymbol{E}}}
\newcommand{\bfpp}{{\boldsymbol{P}}}

%\begin{center}
\title{Supplementary Information: Photocurrent-driven transient symmetry breaking in the Weyl semimetal TaAs}
%\maketitle

%\tableofcontents
%\newpage
\setcounter{section}{0}
\section{Isolated Photoinduced Change in the SHG pattern}

\setcounter{figure}{0}
\begin{figure}[h]
    \centering     
    \includegraphics[width=\columnwidth]{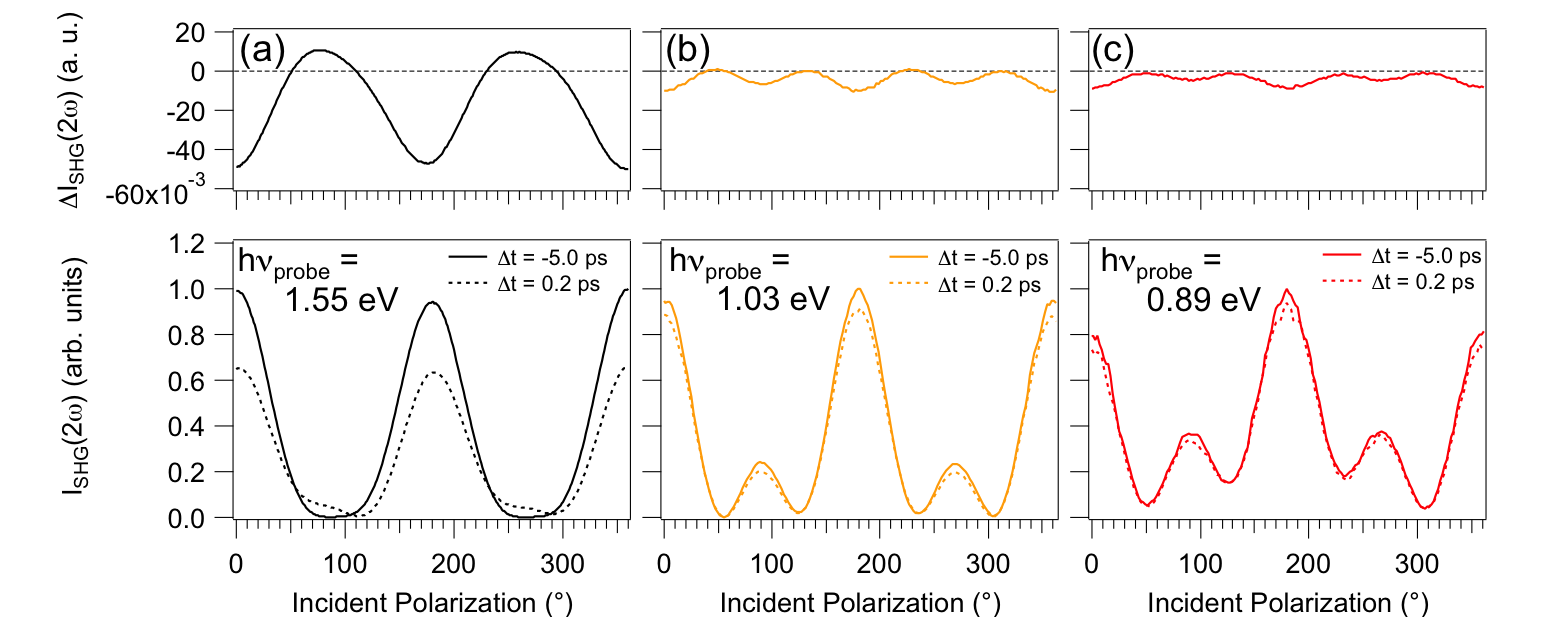}
    \caption{Photoinduced change measured across the entire [1,1,$\bar{1}$] SHG pattern at a fixed time delay (0.2 ps). Here, photoinduced changes are isolated through taking the difference between pre- and post-pump SHG patterns generated from (a) $\hbar\omega = 1.55$ eV, (b) $\hbar\omega = 1.03$ eV and (c) $\hbar\omega = 0.89$ eV probe energies.}
    \label{fig:Supp_PI}
\end{figure}

\newpage
\section{Time-Resolved SHG Traces With IR Probes}

\begin{figure}[h]
    \centering     
    \includegraphics[width=\columnwidth]{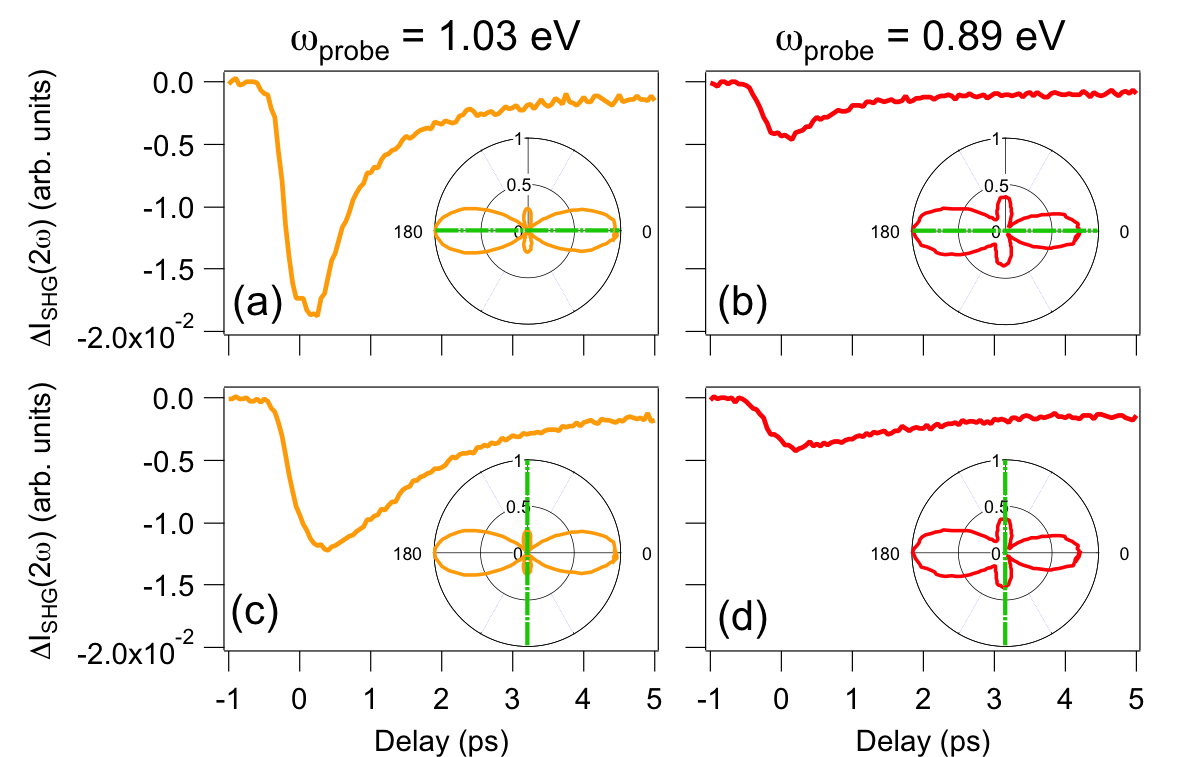}
    \caption{Time-dependent traces of $\Delta I_{\text{SHG}}(2\omega)$ measured for (a,c) 1.03 eV and (b,d) 0.89 eV probe energies following a 1.55 eV pump excitation (fluence = 4.34 mJ/cm$^{2}$). Traces reveal a suppression of both the (a,b) main lobe ($0^{\circ}$) and (c,d) the minor lobe ($90^{\circ}$) for the [1,1,$\bar{1}$] SHG patterns shown as insets.}
    \label{fig:Traces_IR}
\end{figure}

\newpage
\section{Time-resolved X-ray diffraction}

\begin{figure}[b]
    \centering     
    \includegraphics[width=0.63\columnwidth]{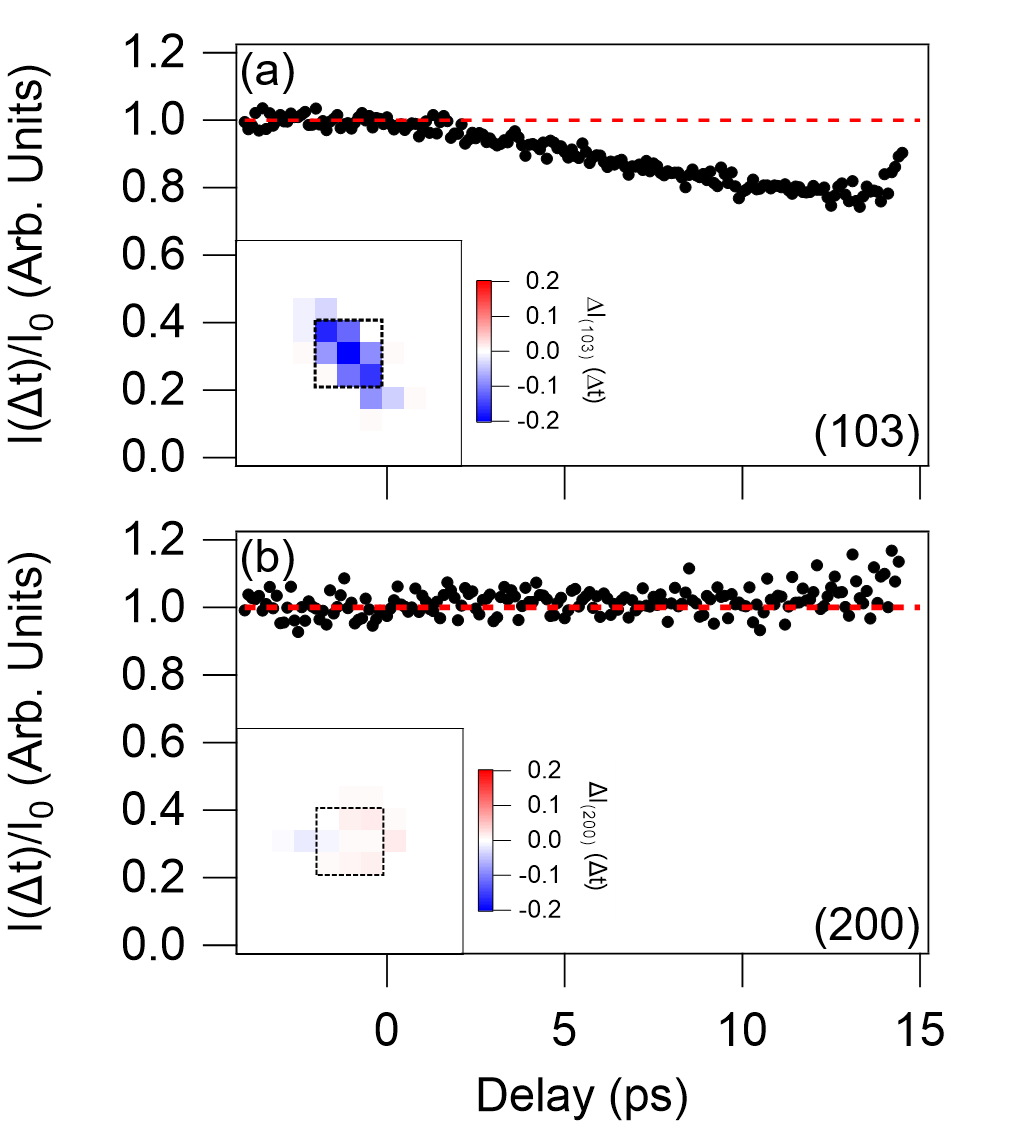}
    \caption{Lattice dynamics of the (a) (103) and (b) (200) Bragg reflections, as integrated over the $3\times3$ pixel area ($\sim4.0\times10^{-2}~^{\circ}$/pixel) shown in the inset.}
    \label{fig:trXRD_Bragg}
\end{figure}

  Time-resolved X-ray diffraction (TR-XRD) experiments were performed on a TaAs single crystal having a surface normal along the (112) direction. This experiment was carried out on the X-ray pump-probe (XPP) instrument \cite{Chollet:yi5004} at the Linac Coherent Light Source (LCLS) \cite{Bostedt2016}. Optical excitation from an amplified Ti:Sapphire (1.55 eV) laser system operating at a 120 Hz repetition rate was chosen to closely match the experimental conditions used in our TR-SHG study. Lattice dynamics probed by a 35 fs, monochromatic X-ray pulse centered at 9.52 keV were measured following photoexcitation by an optical pump pulse having an excitation fluence of 2.86 mJ/cm$^{2}$. Experiments were performed in a reflection geometry, with the X-ray probe having a grazing angle of 0.5$^\circ$ with respect to the (112) face, and chosen to closely match the penetration depth of a normally incident optical pump pulse. Shot-to-shot fluctuations in the time delay between the optical pump and X-ray probe were corrected for by a time diagnostic tool \cite{Harmand2013}, leading to a temporal resolution better than 80 fs. 

  Lattice dynamics of the (103) and (200) Bragg reflections, allowed by the tetragonal symmetry of TaAs, are shown in Fig. S3. Here, a 20\% attenuation of the (103) Bragg peak following 1.55 eV pump excitation occurs over a $\sim 10$ ps timescale, consistent with lattice heating captured by the Debye-Waller effect. No such attenuation is observed for the (200) Bragg reflection, due to a dependence of the Debye-Waller factor on the scattering vector, $\vec{q}$ \cite{WarrenXRD}. The insets in Fig. S3(a-b) depict a change in intensity of the (103) and (200) Bragg reflections, as defined by $\Delta I = I(\Delta t = 10 ~\text{ps})-I(\Delta t = -2 ~\text{ps})$, revealing that the position and structure factor of these Bragg peaks remains constant over short timescales, suggesting the lattice plays a secondary role. In conjunction with the fact that SHG patterns measured with probe energies different from the 1.55 eV excitation energy retain 4$mm$1' symmetry, these TR-XRD findings further emphasize that structural dynamics cannot underlie the symmetry breaking observed in the SHG pattern when resonantly probing the transiently excited state.
  %Hence, the tetragonal symmetry of the lattice, as probed by symmetry allowed Bragg reflections, is preserved over all time delays following optical excitation. This finding is in agreement with the fact that SHG patterns measured with probe energies different from the 1.55 eV excitation energy retain 4mm1' symmetry, further emphasizing that symmetry breaking in the SHG pattern is only observed when .
 
\newpage
\section{Pump Helicity-Dependence of Photoinduced SHG Pattern}

\begin{figure}[h]
    \centering     
    \includegraphics[width=\columnwidth]{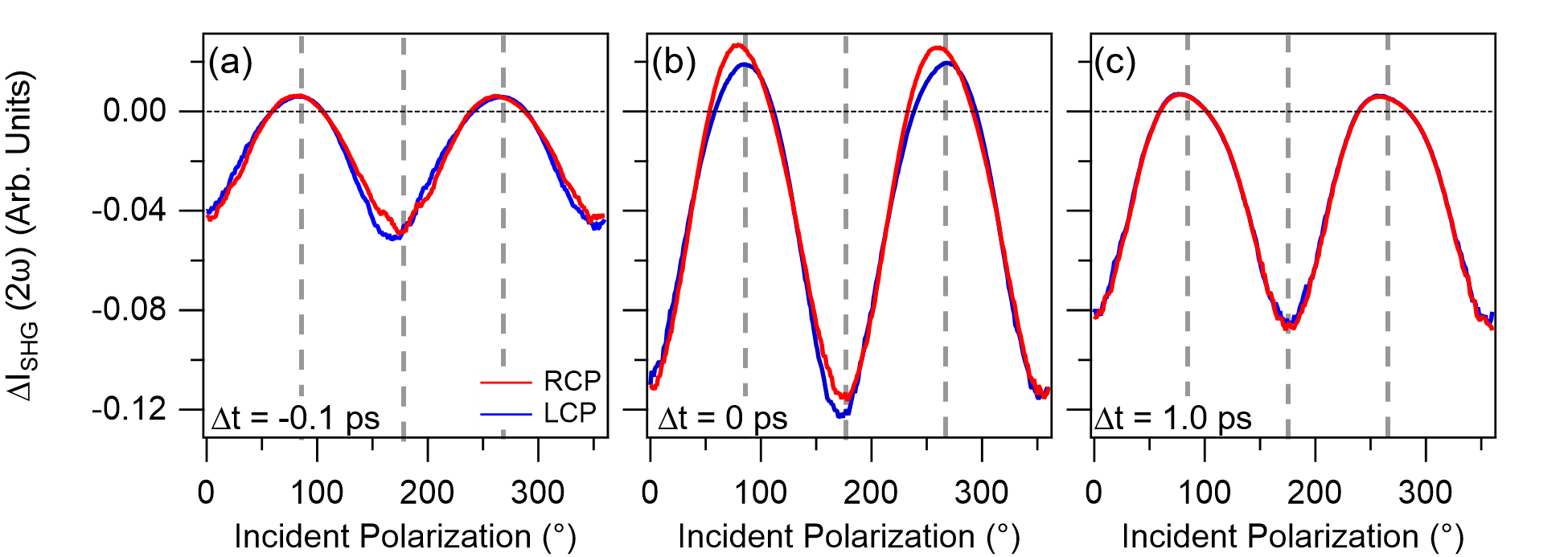}
    \caption{Photoinduced changes to the [1,1,$\bar{1}$] SHG pattern at 1.55 eV measured as a function of pump helicity (right circularly polarized (RCP) vs. left circularly polarized (LCP)) and delay. Dashed lines at $90^{\circ}$, $180^{\circ}$, and $270^{\circ}$ reveal a helicity dependence in the emergent asymmetric lobes as well as in the rotation of the pattern over an ultrashort (a) -0.1 ps and (b) 0 ps timescale, which is lost following a (c) 1.0 ps pump delay.}
    \label{fig:Temp_Dep}
\end{figure}

\newpage
\section{Non-degenerate TR-SHG Traces}

\begin{figure}[h]
    \centering     
    \includegraphics[width=\columnwidth]{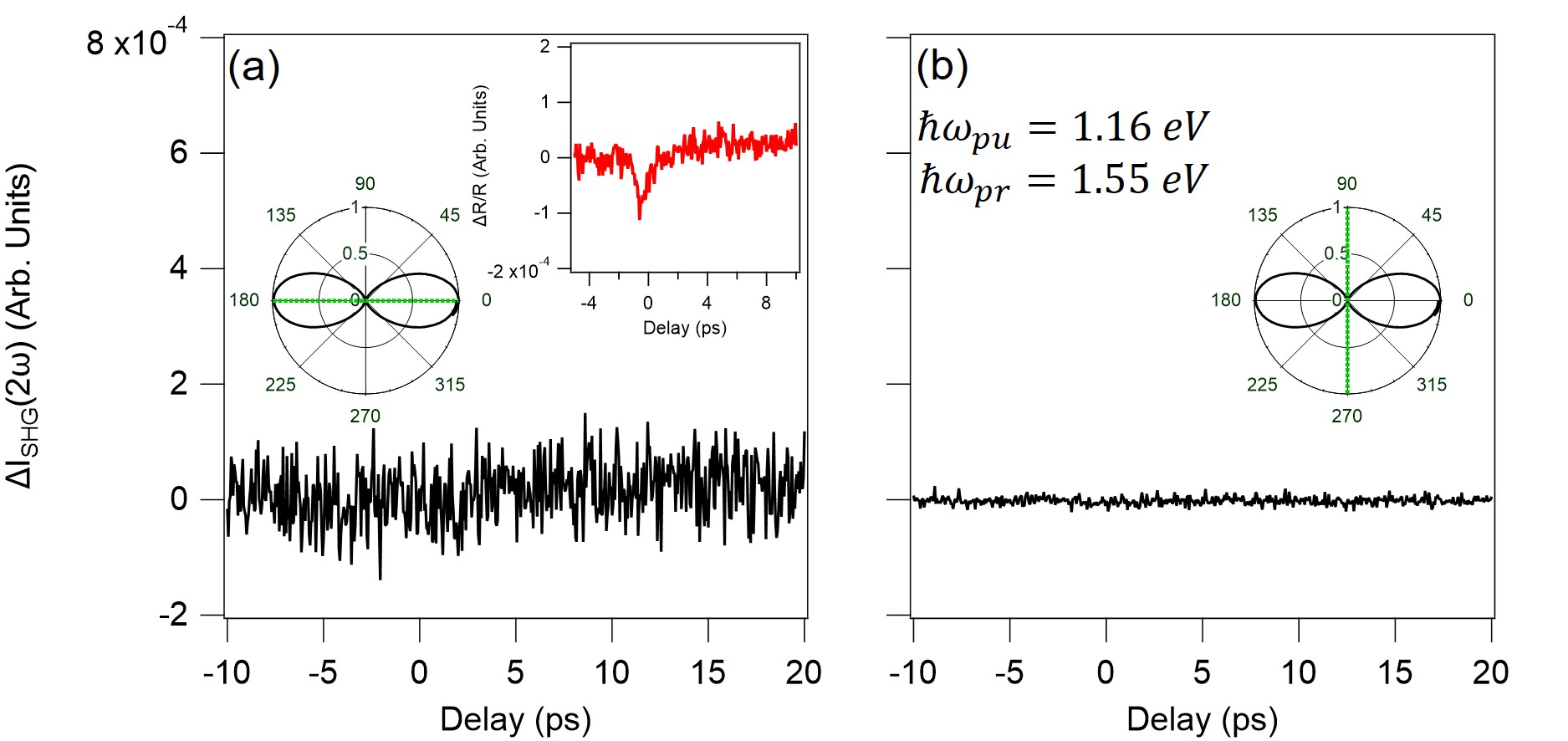}
    \caption{TR-SHG traces measured for input probe polarizations along the green dashed line of the [1,1,$\Bar{1}$] SHG pattern (insets) obtained for $\hbar\omega_{probe} = 1.55$ eV under nondegenerate pump excitation ($\hbar\omega_{pump} = 1.16$ eV; fluence $< 1.4$ mJ/cm$^{2}$). No observable change in the SHG pattern is seen for input polarizations along the (a) [1,1,$\Bar{1}$] and (b) [1,$\Bar{1}$,0] axes. The transient reflectivity shown in the right inset of (a) is consistent with the fact that this experiment does not require the generation of a coherent photocurrent, or any accompanying symmetry change, as it predominantly probes the lifetime of incoherently scattered carriers. In contrast, TR-SHG in TaAs is sensitive to changes in symmetry brought on by the coherent generation of a photocurrent. By probing at a photon energy considerably higher than that used for pump excitation, we ensure that only incoherently scattered carriers can be measured, explaining why a transient reflectivity, and no TR-SHG, response is observed.}
    \label{fig:NonDegn}
\end{figure}

\newpage
\section{Photoinduced Change of the [1,1,$\Bar{1}$] SHG pattern for Degenerate 1.02 eV Photoexcitation}

\begin{figure}[h]
    \centering     
    \includegraphics[width=\columnwidth]{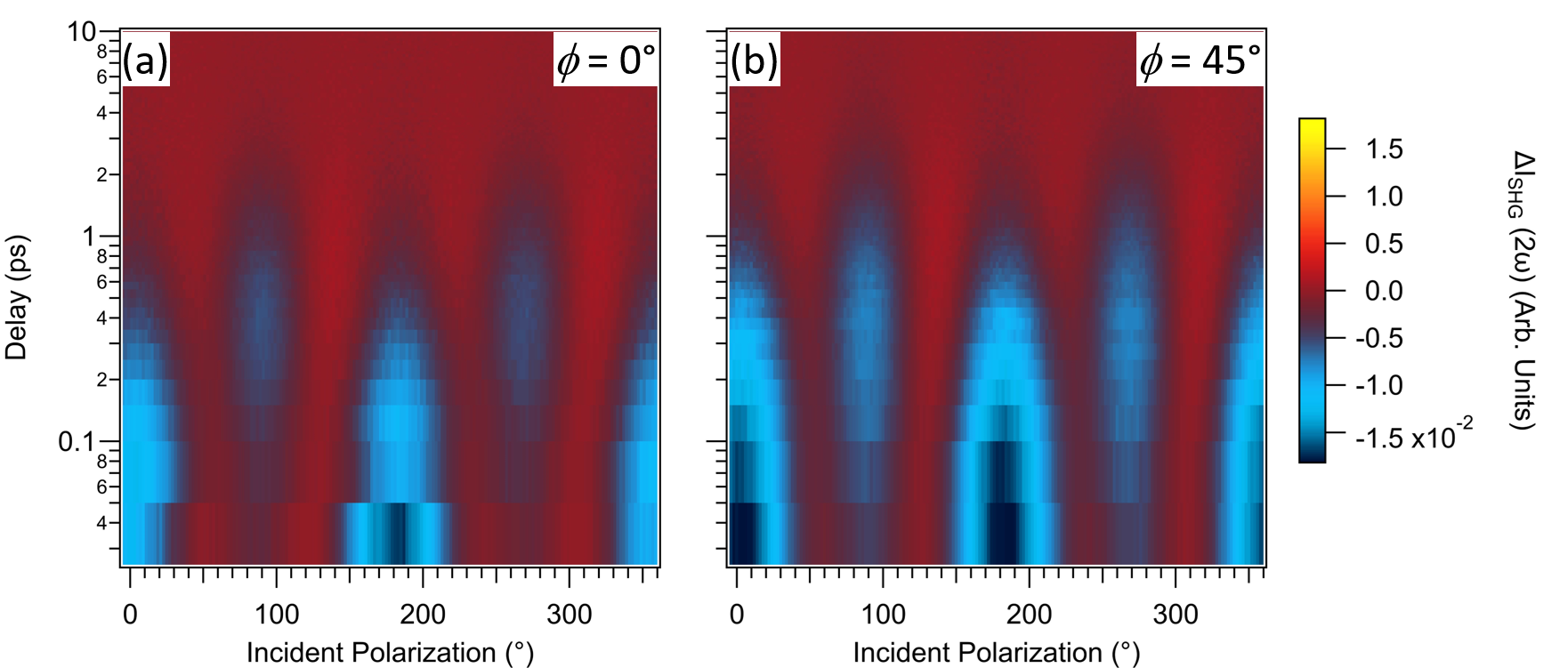}
    \caption{Photoinduced polarization- and time-dependent changes in SHG intensity, $\Delta$I$_{SHG}$(2$\omega$), measured across the entire [1,1,$\Bar{1}$] pattern following degenerate pump and probe excitation at 1.02 eV (fluence $< 1.4$ mJ/cm$^{2}$). While a polarization dependence in the dynamics is observed for pump excitations polarized either (a) parallel ($0^{\circ}$) or at (b) $45^{\circ}$ with respect to the [1,1,$\Bar{1}$] axis, a suppression of the SHG response dominates, with the pattern itself retaining $4mm1'$ symmetry within our experimental resolution. By changing the photon energy, both the photocurrent and SHG response can change in magnitude, but this does not necessarily translate to a stronger degree of transient symmetry breaking, as it is more difficult to detect photocurrent-induced changes along [1,-1,0] for a given input polarization when static SHG dominates the nonlinear response. %, especially in cases where $\chi_{xxz}(\omega) > \sigma_{xxz}(\omega)$ as occurs for $\hbar\omega < 1.55$ eV. 
    Instead, the pronounced anisotropy in the SHG pattern at this particular photon energy is what allows for the clear observation of photocurrent-induced SHG, leading to the generation of a transient $1$ symmetry state.}
    \label{fig:PI_1eV}
\end{figure}

\newpage
\section{Fluence Dependence}

\begin{figure}[h]
    \centering     
    \includegraphics[width=\columnwidth]{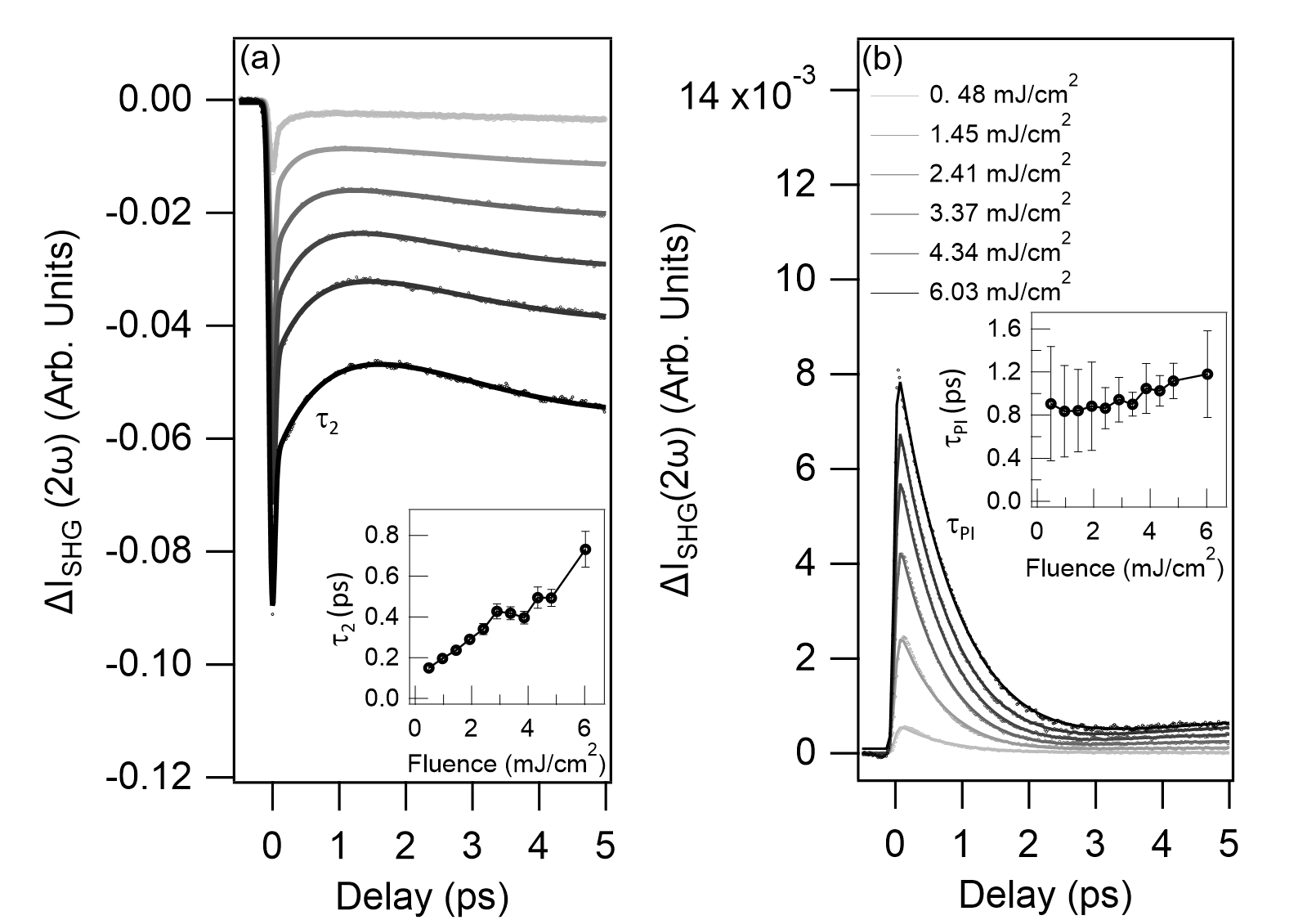}
    \caption{Room temperature, pump fluence dependence of the time-resolved SHG traces at 1.55 eV for (a) the main lobe ($0^{\circ}$) and (b) the asymmetric, photoinduced lobe ($90^{\circ}$) present in the [1,1,$\Bar{1}$] SHG pattern. The inset shows the fluence dependence of the relaxation times $\tau_{2}$ and $\tau_{\text{PI}}$, as determined from fits following linearly polarized pump excitation. Here, error bars denote one standard deviation}
    \label{fig:Fluence_Dep}
\end{figure}

\newpage
\section{Temperature Dependence}

\begin{figure}[h]
    \centering     
    \includegraphics[width=\columnwidth]{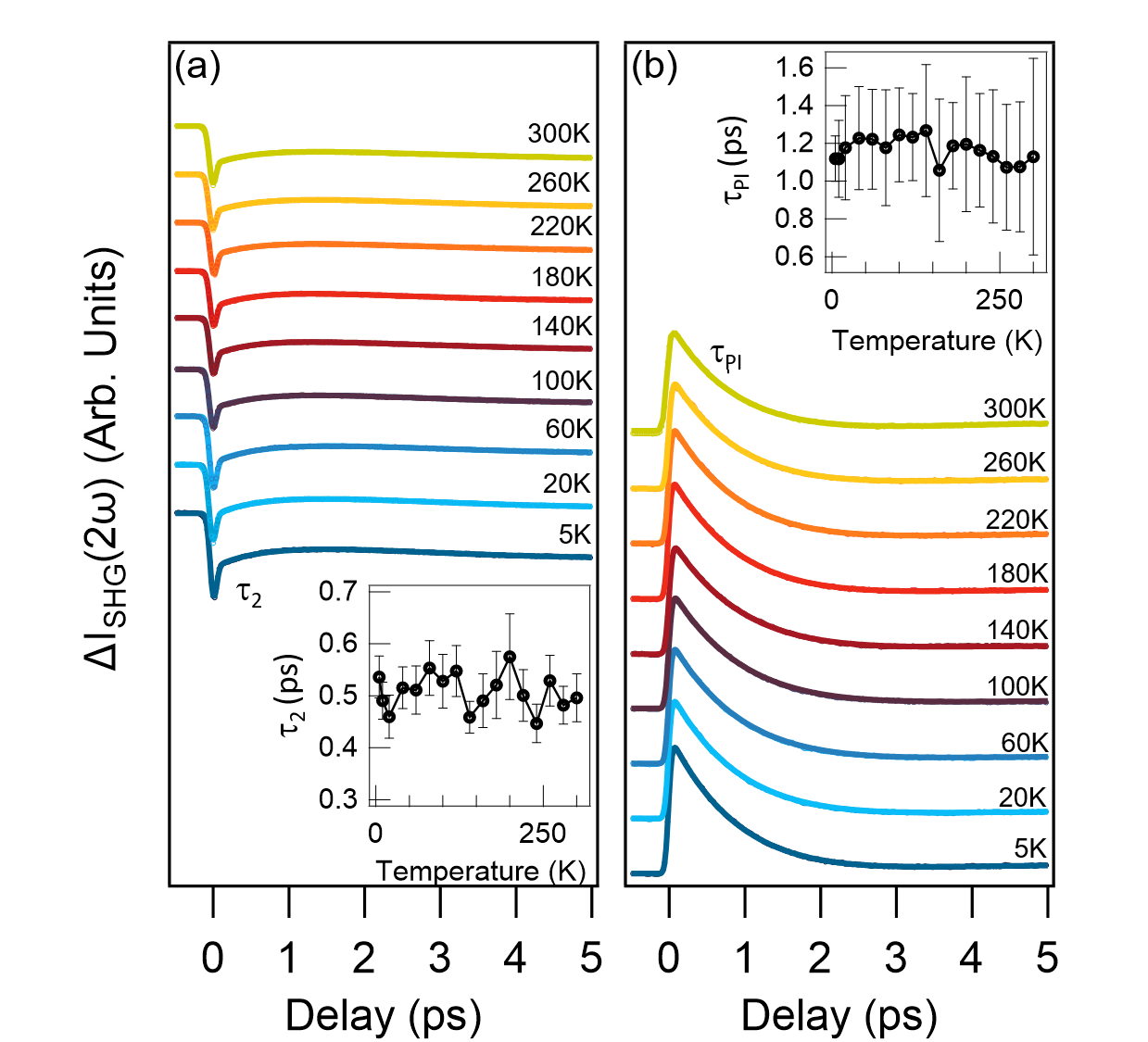}
    \caption{Temperature dependence of the time-resolved SHG traces at 1.55 eV for (a) the main lobe ($0^{\circ}$) and (b) the asymmetric, photoinduced lobe ($90^{\circ}$) present in the [1,1,$\Bar{1}$] SHG pattern. The inset shows the temperature dependence of the relaxation time $\tau_{2}$ and $\tau_{\text{PI}}$ (main text) as determined from fits following linearly polarized pump excitation with a fluence of 4.34 mJ/cm$^{2}$. Here, error bars denote one standard deviation}
    \label{fig:Temp_Dep}
\end{figure}

\newpage
\section{Pump Induced Shift Currents in TaAs}

The shift current is defined by Sipe and Shkrebtii \cite{Sipe2000} in terms of the nonlinear conductivity tensor $\sigma_{abc}$ as
\begin{align*}
    J_{a}&=\sum_{b,c=x,y,z}\sigma_{abc}E_{b}(\omega)E_{c}(-\omega),
\end{align*}
where the electric field of the incident light pulse is given by $E(t)=E(\omega)e^{-i \omega t}+E(-\omega)e^{i \omega t}$. Based on symmetry, the 4mm1' point group of TaAs constrains $\sigma_{abc}$ to yield three independent tensor elements: $\sigma_{xxz}=\sigma_{yyz}=\sigma_{xzx}=\sigma_{yzy}$, $\sigma_{zxx}=\sigma_{zyy}$, and $\sigma_{zzz}$. For light incident on the surface normal of the (112) face, the electric field can be expressed in terms of the in-plane [1,1,$\bar{1}$] and [1,$\bar{1}$,0] axes as
\begin{align*} 
    E_{x}&=\frac{a}{|a_{1}'|}E_{[1,1,\bar{1}]}+\frac{a}{|a_{2}'|}E_{[1,\bar{1},0]}\\
    E_{y}&=\frac{a}{|a_{1}'|}E_{[1,1,\bar{1}]}-\frac{a}{|a_{2}'|}E_{[1,\bar{1},0]}\\
    E_{z}&=-\frac{c}{|a_{1}'|}E_{[1,1,\bar{1}]},
\end{align*}
where $|a_{1}'|=\sqrt{2}a\sqrt{1+\frac{c^2}{2a^2}}$ and $|a_{2}'|=\sqrt{2}a$ define normalization constants in terms of the lattice parameters $a$ and $c$. Hence, the excitation of a shift current along the [1,$\bar{1}$,0] axis is allowed by symmetry under the condition 
\begin{align*} 
    J_{[1,\bar{1},0]}&=\frac{a}{|a_{2}'|}(J_{x}-J_{y})\\
    &=\frac{a}{|a_{2}'|}(\sigma_{xxz}E_{x}(\omega)E_{z}(-\omega)+\sigma_{xzx}E_{z}(\omega)E_{x}(-\omega)-\sigma_{yyz}E_{y}(\omega)E_{z}(-\omega)-\sigma_{yzy}E_{z}(\omega)E_{y}(-\omega))\\
    &=-\frac{1}{\sqrt{1+2(\frac{a}{c})^2}}\sigma_{xxz}(E_{[1,1,\bar{1}]}(\omega)E_{[1,\bar{1},0]}(-\omega)+E_{[1,1,\bar{1}]}(-\omega)E_{[1,\bar{1},0]}(\omega)),
\end{align*}
while shift current generation along [1,1,$\bar{1}$] follows from 
\begin{align*} 
    J_{[1,1,\bar{1}]}&=\frac{1}{|a_{1}'|}(aJ_{x}+aJ_{y}-cJ_{z})\\
    &=-\frac{1}{(2a^{2}+c^{2})\sqrt{1+2(\frac{a}{c})^{2}}}((4a^{2}\sigma_{xxz}+2a^{2}\sigma_{zxx}+c^{2}\sigma_{zzz})E_{[1,1,\bar{1}]}(\omega)E_{[1,1,\bar{1}]}(-\omega)+\\& ~~~~~~~~~~~~~~~~~~(2a^{2}+c^{2})\sigma_{zxx}E_{[1,\bar{1},0]}(\omega)E_{[1,\bar{1},0]}(-\omega)).
\end{align*}
Thus, shift current generation along [1,1,$\bar{1}$] will always be allowed by symmetry on the (112) face, regardless of polarization. In contrast, the excitation of a shift current along [1,$\bar{1}$,0] requires the polarization to be detuned from either the [1,1,$\bar{1}$] or [1,$\bar{1}$,0] axes, with the strongest contribution coming from an equal projection along these two orthogonal axes (i.e. $45^{\circ}$). Such a result is similar to the symmetry constraints imposed on injection photocurrents, which are restricted to flow along the [1,$\bar{1}$,0] axis only \cite{Sirica_THz_2019}.      
\onecolumngrid
\newpage
\section{Symmetry analysis of time-resolved second-harmonic generation patterns}
\label{sec_SI:symmetry_analyis}
  In this section, we describe our procedure for fitting the SHG patterns obtained before and after pump excitation ($\Delta t = \{-5.0, 0.0, 1.0, 5.0 \}$~ps). In Sec.~\ref{subsec_SI:exp_geometry}, we give details about the experimental setup. Then, in Sec.~\ref{subsec:SHG_tensor_MSG}, we derive the form of the SHG electric dipole tensor for the relevant magnetic point groups (MPGs), state the general form of the expressions for the outgoing intensities in the $[1\bar{1}0]$ and $[11\bar{1}]$ channels, and describe details of our fits.

  \subsection{Experimental setup}
  \label{subsec_SI:exp_geometry}
\begin{figure}[b]
    \includegraphics[width=.4\linewidth]{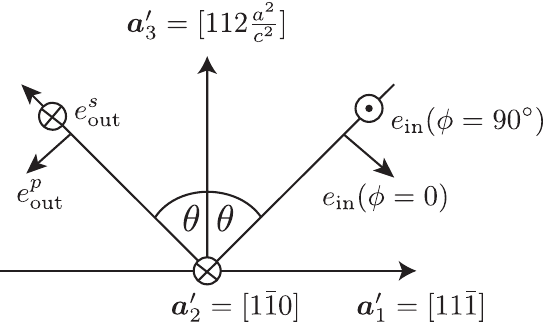}
    \caption{Sketch of the experimental setup in which the incoming probe beam makes an angle, $\theta$, with respect to the surface normal $\bfa'_3=[112\frac{a^2}{c^2}]$ and the in-plane $\bfa'_1 = [11\bar{1}]$ direction. The polarization of the probe beam is continuously rotated over a full $360^\circ$, starting from $\phi = 0^{\circ}$ (p-polarization) parallel to the $\bfa'_1$-$\bfa'_3$ plane.  }
    \label{fig_S7:experimental_setup}
  \end{figure}

The experimental geometry is depicted schematically in Fig.~\ref{fig_S7:experimental_setup}. It shows the scattering of an SHG probe beam relative to the normal of the $(112)$ surface, defined as $\bfa'_3 = \bfa_1 + \bfa_2 + 2 \frac{a^2}{c^2} \bfa_3 \equiv [112 \frac{a^2}{c^2}]$. The two high-symmetry directions on the (112) surface plane are defined as $\bfa'_1 = \bfa_1 + \bfa_2 - \bfa_3 = [11\bar{1}]$ and $\bfa'_2 = \bfa_1 - \bfa_2 = [1\bar{1}0]$. Here, we denote the conventional tetragonal basis vectors as $\bfa_1 \equiv [100]$, $\bfa_2 \equiv [010]$ and $\bfa_3 = [001]$, where $|\bfa_1| = |\bfa_2| = a = 3.4348~\text{\AA}$ and $|\bfa_3| = c = 11.641~\text{\AA}$. We note that the vector $\bfa'_2$ is orthogonal to the polar axis, $[001]$. A transformation from the primed basis vectors to the conventional tetragonal basis vectors is obtained via  $\bfa'_\alpha = \sum_\beta U_{\beta \alpha} \bfa_\beta$ with the transformation matrix
    \begin{equation}
        (U_{\beta \alpha}) = \begin{pmatrix} 1 & 1 & 1 \\ 1 & -1 & 1 \\ -1 & 0 & 2 a^2/c^2 \end{pmatrix}\,. 
    \end{equation}
While the primed lattice vectors $\bfa'_\alpha$ are orthogonal $\bfa'_\alpha \cdot \bfa'_\beta \propto \delta_{\alpha \beta}$, they are not normalized. The length of the primed basis vectors is $|\bfa'_1| = \sqrt{2} a \sqrt{1 + \frac{c^2}{2 a^2}}$, $|\bfa'_2| = \sqrt{2} a$, and $|\bfa'_3| = \sqrt{2} a \sqrt{1 + 2 \frac{a^2}{c^2}}$. It is convenient to introduce normalized basis vectors via $\bfe_\alpha =\bfa_\alpha/|\bfa_\alpha|$ and $\bfe'_\alpha = \bfa'_\alpha/|\bfa'_\alpha|$. The basis transformation matrix between these two orthonormal basis sets is achieved via $\bfe'_\alpha = \sum_\beta \tilde{U}_{\beta \alpha} \bfe_\beta$ with transformation matrix
\begin{equation}
(\tilde{U}_{\beta \alpha}) = \begin{pmatrix} a/|\bfa'_1| & a/|\bfa'_2| & a/|\bfa'_3| \\ a/|\bfa'_1| & -a/|\bfa'_2| & a/|\bfa'_3| \\ - c/|\bfa'_1|  & 0 & 2 a^2/(c |\bfa'_3|)
\label{eq:S0_transformation_matrix}
\end{pmatrix} \,.
    %\tilde{U}_{\beta \alpha} = \begin{pmatrix} 1/\mathcal{N\textbf{}}_1 & 1/\mathcal{N}_2 & 1/\mathcal{N}_3 \\ 1/\mathcal{N}_1 & -1/\mathcal{N}_2 & 1/\mathcal{N}_3 \\ - c/a \mathcal{N}_1 & 0 & 2 a/c \mathcal{N}_3 \end{pmatrix}
\end{equation}
 Expressed in this orthonormal basis, the components of a vector transform according to $\bfv = \sum_\alpha v'_\alpha \bfe'_\alpha = \sum_\beta \Bigl( \sum_\alpha \tilde{U}_{\beta\alpha} v'_\alpha \Bigr) \bfe_\beta = \sum_\beta v_\beta \bfe\textbf{}_\beta$, leading to 
  $v_\beta = \sum_{\alpha} \tilde{U}_{\beta \alpha} v'_\alpha$. 

  A $\tau_{\text{pump}} = 80$~fs pump pulse centered at a wavelength $\lambda=800$~nm is directed along the normal $\bfa'_3 = [112\frac{a^2}{c^2}]$ axis of the crystal surface. The pump pulse is linearly polarized along $\hat{e}_\text{pump} = \bfe'_1 (\propto [11\bar{1}])$, up to about a 2$^\circ$ alignment error. The incoming probe beam has a variable wavelength between $\lambda=800$~nm and $\lambda=1400$~nm, making an angle of $\theta = 6^\circ$ with respect to the surface normal $\bfa'_3=[112\frac{a^2}{c^2}]$. The scattering plane of the probe beam is defined by the $\bfe'_1$-$\bfe'_3$ plane as given by the incoming and outgoing wavevectors
  \begin{align}
    \label{eq:S1}
    \bfk_{\text{in/out}} &= \frac{2 \pi}{\lambda} \Bigl( - \sin \theta \, \bfe'_1 \mp \cos \theta \,\bfe'_3 \Bigr) \,,
%\label{eq:S2}
%\bfk_{\text{out}} &= \frac{2 \pi}{\lambda} \Bigl(- \sin \theta \hat{\bfx}' + \cos \theta \hat{\bfz}' \Bigr) \,.
  \end{align}
  where the upper sign refers to $\bfk_{\text{in}}$. In the experiment, the incoming polarization, $\hat{\bfe}_\text{in}(\phi)$, is continuously rotated, and in the primed coordinate system takes the form 
  \begin{equation}
    \label{eq:S3}
    \hat{\bfe}_\text{in}(\phi) = R(\phi, \hat{\bfk}_{\text{in}}) (\cos \theta, 0, - \sin\theta)^T \,.
  \end{equation}
  Here, the rotation matrix $R(\phi, \hat{\bfk}_{\text{in}})$ describes a rotation by angle $\phi$ around the direction $\hat{\bfk}_{\text{in}}= \bfk_{\text{in}}/|\bfk_{\text{in}}|$. Explicitly, $\hat{\bfe}_{\text{in}}(\phi = 0^\circ) = \cos \theta \bfe'_1 - \sin \theta \bfe'_3$ corresponds to $p$-polarization and $\hat{\bfe}_{\text{in}}(\phi = 90^\circ) = -\bfe'_2$ corresponds to $s$-polarization. 

  As shown in Fig.~\ref{fig_S7:experimental_setup}, we record the outgoing SHG intensities in two channels: one parallel to $\bfa'_2 = [1\bar{1}0]$ (s-out) and one primarily parallel to $\bfa'_1 = [11\bar{1}]$ (p-out):
  \begin{align}
    \label{eq:S4}
    I_{\text{SHG}}^{[1\bar{1}0]}(2 \omega; \phi) &\propto |\bfpp(2 \omega; \phi) \cdot \bfe'_2|^2 \\
    \label{eq:S5}
    I_{\text{SHG}}^{[11\bar{1}]}(2 \omega; \phi) &\propto |\bfpp(2 \omega; \phi) \cdot \bfe'_1|^2 \,.
  \end{align}
  Here, $\bfpp(2 \omega; \phi)$ is the nonlinear SHG polarization that is induced in the material at twice the frequency of the incoming light $\omega = 2 \pi c/\lambda$. Since TaAs is non-centrosymmetric, i.e., it lacks inversion symmetry, the polarization is dominated by the bulk electric dipole response 
  \begin{equation}
    \label{eq:S6}
    \bfpp_i(2 \omega; \phi) = \sum_{j,k = x,y,z} \chi^{\text{ED}}_{ijk}(2 \omega; \omega, \omega) \bfee_j(\omega; \phi) \bfee_k(\omega; \phi)\,,
  \end{equation}
  where the incoming electric field is given by $\bfee(\omega; \phi) = E(\omega) \hat{\bfe}_{\text{in}}(\phi)$, and the indices $i,j,k \in \{x,y,z\}$ refer to the conventional tetragonal basis directions $\bfa_1$, $\bfa_2$, $\bfa_3$. 

  \subsection{SHG electric dipole response tensor for relevant magnetic point groups}
  \label{subsec:SHG_tensor_MSG}
  In general, the non-linear electric dipole susceptibility $\chi^{\text{ED}}_{ijk}$ has $18$ independent complex elements due to permutation symmetry in the last two indices (see Eq.~\eqref{eq:S6}):
  %\begin{widetext}
  \begin{align}
    \chi_{ijk}^{\text{ED}} \equiv \chi_{ijk} = \begin{pmatrix}  \begin{pmatrix} xxx & xxy & xxz \\ xxy & xyy & xyz \\ xxz & xyz & xzz \end{pmatrix} \begin{pmatrix} yxx & yxy & yxz \\ yxy & yyy & yyz \\ yxz & yyz & yzz \end{pmatrix} \begin{pmatrix} zxx & zxy & zxz \\ zxy & zyy & zyz \\ zxz & zyz & zzz \end{pmatrix}\end{pmatrix} \,, 
    \label{eq:chi_ED_general}
  \end{align}
  %\end{widetext}
  \onecolumngrid
  where we will drop the superscript ``ED''for the remainder of this discussion.  
  The presence of magnetic point symmetries in the crystal puts constraints on the form of the tensor, according to Neumann's principle. Under a transformation with an element $R$ of the point group, represented by a matrix $R_{ij}$, one finds
  \begin{equation}
    \chi_{ijk} \xrightarrow[]{R} \tilde{\chi}_{ijk} = \sum_{i', j', k'} R_{ii'} R_{jj'} R_{kk'} \chi_{i'j'k'}
  \end{equation}
  where $\tilde{\chi}_{ijk} = \chi_{ijk}$ by symmetry. Under time-reversal $\mathcal{T}$, which acts as complex conjugation $K$, the tensor transforms as $\chi_{ijk} \xrightarrow[]{\mathcal{T}} \tilde{\chi}_{ijk} = \chi_{ijk}^*$. In the presence of time-reversal symmetry, all elements of $\chi_{ijk}$ are thus real. For an element of the MPG that combines a spatial symmetry with time-reversal $R' = R \mathcal{T}$, one finds the constraint $\chi_{ijk} \xrightarrow[]{R' = \mathcal{T}R} \tilde{\chi}_{ijk} = \sum_{i', j', k'} R_{ii'} R_{jj'} R_{kk'} \chi^*_{i'j'k'} = \chi_{ijk}$. %Below, we explicitly state the form of $\chi_{ijk}$ for the different (magnetic) point group symmetries relevant in our experiment. 

  \onecolumngrid
  \subsubsection{4mm1' symmetry before and long after pump excitation}
  TaAs is characterized by the crystalline point group $C_{4v}=4\text{$mm$}$ and possesses time-reversal symmetry in the absence of any photocurrent. The relevant MPG before and long ($>2.0$ ps) after the pump pulse is therefore 4$mm$1'. The point group 4$mm$ consists of a fourfold rotation $4_{0,0,z}$ around the polar $\bfa_3 = [001]$ axis and four vertical mirror planes that contain the polar axis. Two mirror planes are along the tetragonal coordinate axes $m_{x,0,z}$ and $m_{0,y,z}$ and two are along the diagonals $m_{x,x,z}$ and $m_{x,-x,z}$. Note that $x,y,z$ refer to $\bfa_1=[100]$, $\bfa_2 = [010]$ and $\bfa_3=[001]$. It is thus convenient to work in the unprimed (crystal) basis and express the electric field $\bfee(\omega)$ in Eq.~\eqref{eq:S6} in this basis using the transformation matrix~\eqref{eq:S0_transformation_matrix}.

  For 4$mm$1' symmetry, the nonlinear susceptibility contains only three independent real elements, and we can use $\{xxz, zxx, zzz\} \in \mathbb{R}$ to parameterize it. Fourfold rotation symmetry, for example, which is expressed by the matrix $R[4^+_{0,0,z}] = \left( \begin{smallmatrix} 0 & -1 & 0 \\ 1 & 0 & 0 \\ 0 & 0 & 1 \end{smallmatrix} \right)$ and $R[4^-_{0,0,z}] = \left( \begin{smallmatrix} 0 & 1 & 0 \\ -1 & 0 & 0 \\ 0 & 0 & 1 \end{smallmatrix} \right)$, reduces the number of independent elements to four: $\{xxz, xyz, zxx, zzz\}$. The presence of mirror symmetries such as $R[m_{x,0,z}] = \left( \begin{smallmatrix} 1 & 0 & 0 \\ 0 & -1 & 0 \\ 0 & 0 & 1 \end{smallmatrix} \right)$ enforces $xyz \rightarrow 0$, resulting in the form of the nonlinear electric-dipole susceptibility for 4$mm$ symmetry to be
  
  %\begin{widetext}
  \begin{equation}
    \chi_{ijk}^{(C_{4v})} = \begin{pmatrix}\begin{pmatrix} 0 & 0 & xxz \\ 0 & 0 & 0\\ xxz & 0 & 0 \end{pmatrix} \begin{pmatrix} 0 & 0 & 0 \\ 0 & 0 & xxz \\ 0 & xxz & 0 \end{pmatrix} \begin{pmatrix} zxx & 0 & 0 \\ 0 & zxx & 0 \\ 0 & 0 & zzz \end{pmatrix} \end{pmatrix} \,.
    \label{eq:chi_C4v}
  \end{equation}
  %\end{widetext}
  
  \onecolumngrid
  The tensor elements are real if the system is time-reversal symmetric (4$mm$1') and complex if time-reversal is broken (4$mm$). The outgoing intensities in the two channels we measure read
  \begin{align}
    \label{eq:S7}
    I^{[1\bar{1}0]}_{\text{SHG}} &= a_1 \sin^2(2 \phi)\\
    I^{[11\bar{1}]}_{\text{SHG}} &= \bigl[b_1 + b_2 \cos^2(\phi) \bigr]^2 \,.
  \end{align}
  For a fixed incoming angle $\theta$, the coefficient $a_1(xxz)$ is a function of $xxz$ only and $b_1(zxx)$ will depend on $zxx$ only. The coefficient $b_2(xxz, zxx, zzz)$ depends on all three tensor elements. Note that an (unknown) global proportionality factor has been absorbed into this definition for the matrix elements. As a result, fitting our experimental data, which is given in arbitrary units, only yields the ratios of tensor elements, but not their absolute values. 

  The outgoing intensities for the relevant MPG symmetries are collected in Table~\ref{tab:outgoing_I_relevant}. In addition to 4mm1' symmetry, which is relevant for the bulk, static pattern, as well as for long ($>2.0$ ps) pump delays, we also include the form of the outgoing intensities for m1' symmetry with diagonal mirror $m_{x,x,z}$. This is the MPG of the (112) surface, which has only one mirror plane in addition to time-reversal. Interestingly, we find that the general form of the outgoing intensities is identical to the case of 4mm1' symmetry, which is likewise found to be valid in the absence of time-reversal. Adding the intensity of an additional electric-dipole surface response therefore does not allow for the overall rotation and photoinduced asymmetric lobes in the transiently excited $I^{[11\bar{1}]}_{\text{SHG}}$ pattern to be fit.
  \onecolumngrid
  \begin{table}[b!]
  \begin{center}
  \begin{tabular}{c||c|c}
  MPG & Form of $I_\text{SHG}^{\text{[1$\bar{1}$0]}}$ & Form of $I_\text{SHG}^{\text{[11$\bar{1}$]}}$ 
\\
  \hline 
  \hline
  4mm1' &  $a_1 s_{2 \phi}^2$ & $\bigl(b_1 + b_2 c_\phi^2 \bigr)^2 $ \\
  4mm &  $a_1 s_{2 \phi}^2$& $b_1 c_\phi^4 + b_2 s_{\phi}^4 +b_3 s_{2\phi}^2$ \\
  \hline
  m1' $(m_{x,x,z})$&  $a_1 s_{2 \phi}^2$& $\bigl(b_1 + b_2 c_\phi^2 \bigr)^2$ \\
  m $(m_{x,x,z})$ & $a_1 s_{2 \phi}^2$ & $b_1 c_\phi^4 + b_2 s_{\phi}^4 +b_3 s_{2\phi}^2$\\
  \hline
  1' &  $\bigl(  a_1 + a_2 c_\phi^2 + a_3 s_{2\phi}  \bigr)^2$ & $\bigl(b_1 + b_2 c_\phi^2 + b_3 s_{2\phi} \bigr)^2$ \\
  1 & $a_1 c_\phi^4 + a_2 c_\phi^3 s_\phi + a_3 c_\phi s_\phi^3 + a_4 s_\phi^4 + a_5 s_{2 \phi}^2$ & $b_1 c_\phi^4 + b_2 c_\phi^3 s_\phi + b_3 c_\phi s_\phi^3 + b_4 s_\phi^4 + b_5 s_{2\phi}^2$\\
  \hline
% \hline
% \multicolumn{2}{c}{Item} \\
% \cline{1-2}
  \end{tabular}
  \caption{General form of the outgoing intensities along $[1\Bar{1}0]$ (s-out) and $[11\Bar{1}]$ (p-out) for the different MPG symmetries occurring in our experiment. Here, $c_\phi \equiv \cos \phi$, $s_\phi \equiv \sin \phi$ and the coefficients $a_i$ and $b_i$ are real. Before and long after ($\Delta t = \mp 5$~ps) pump excitation, the system has $4mm1'$ symmetry. In the presence of a pump-induced photocurrent, all spatial symmetries and time-reversal symmetry are lost, leaving the system in a reduced $1$ symmetry state. We note that in the main text, we use the following notation for $1$ symmetry: $I_{\text{SHG}}^{[11\Bar{1}]} = \sum_{n=0}^4 \mathcal{C}_n^{[11\Bar{1}]} \sin^n(\phi) \cos^{4-n}(\phi)$, corresponding to $\mathcal{C}^{[11\Bar{1}]}_0 = b_1$, $\mathcal{C}^{[11\Bar{1}]}_1 = b_2$, $\mathcal{C}^{[11\Bar{1}]}_2 = 2 b_5$, $\mathcal{C}^{[11\Bar{1}]}_3 = b_3$, and $\mathcal{C}^{[11\Bar{1}]}_4 = b_4$. Since the symmetry of the $(112)$ surface is given by $m1'$, where the diagonal mirror $m_{x,x,z}$ is preserved, we also include the form of the outgoing intensities for the $m1'$ point group. Because the form of the expression for $4mm1'$ and $m1'$ with diagonal mirror $m_{x,x,z}$ are identical, we conclude that the overall rotation and asymmetric lobes present at $\phi=90^{\circ}$ and $\phi=270^{\circ}$ in the photoinduced $[11\Bar{1}]$ pattern, cannot be reproduced by considering an incoherent surface contribution.}\label{tab:outgoing_I_relevant}
  \end{center}
  \end{table}
  \onecolumngrid
  The ratio of the fit parameters resulting from the best fits at $\Delta t = \mp 5.0$~ps in Table~\ref{tab:4mm_fit_values} show agreement with previous studies~\cite{Patankar_SHG_2018}, where the $zzz$ element is similarly found to be  larger than the other two. The table also includes $R^2$ values of our fits from either the [1$\bar{1}$0] or [11$\bar{1}$] output channels. We find $R^2_{[1\bar{1}0]} = 0.97$ and $R^2_{[11\bar{1}]} = 0.99$, demonstrating that the fit accurately captures our experimental data. 
  \onecolumngrid
  \begin{table}[h!]
  \begin{center}
  \begin{tabular}{c||c|c}
  4mm1' & $\Delta t = -5.0$~ps & $\Delta t = 5.0$~ps
\\
  \hline 
  \hline
  $zxx/xxz$ & $0.12$ & $-0.099$\\
  $zzz/xxz$ & $7.4$ & $5.4$ \\
  \hline
  $b_1/\sqrt{a_1}$ &$0.11$ & $-0.091$\\
  $b_2/\sqrt{a_1}$ & $6.4$& $4.9$\\
  \hline \hline
  $R^2([1 \bar{1}0])$ & $0.97$& $ 0.97$ \\ 
  $R^2([11\bar{1}])$ & $0.99$ & $0.99$  \\
  \hline
% \hline
% \multicolumn{2}{c}{Item} \\
% \cline{1-2}
  \end{tabular}
  \caption{Values of fit parameters and corresponding $R^2$ values for the best fits at $\Delta t = \mp 5.0$~ps for a 4mm1' symmetric tensor $\chi_{ijk}$. The fits are shown in Fig.~1 of the main text in panels (a,e) and (d,f). We observe that the $zzz$ element dominates as expected.}\label{tab:4mm_fit_values}
  \end{center}
  \end{table}
  \onecolumngrid
  \subsubsection{1 symmetry in the transiently excited state}
  \label{subsec:1_symmetry_post_pump}
  A linearly polarized pump pulse normally incident on the surface will induce a transient photocurrent directed at $\sim6^\circ$ relative to the $m_{x,x,z}$ mirror plane of the $(112)$ surface~\cite{Sirica_THz_2019}. This photocurrent will break all spatial symmetries along with time-reversal symmetry. The MPG in the transiently excited state is therefore $1$. The photocurrent decays on the timescale of $\tau_{\text{PI}}\sim 1.1~\text{ps}$ after which 4mm1' symmetry is restored (see panels (d,h) in Fig.~1. and Table~\ref{tab:4mm_fit_values}).

  The nonlinear electric dipole tensor for $1$ symmetry is of the general form in Eq.~\eqref{eq:chi_ED_general} with complex elements. The outgoing intensities $I^{[1\bar{1}0]}_{\text{SHG}}$ and $I^{[11\bar{1}]}_{\text{SHG}}$ are then described by polynomials in $\cos \phi$ and $\sin \phi$ given in Table~\ref{tab:outgoing_I_relevant}. The coefficients $a_{i}$ and $b_i$ for $i=1,\ldots 5$ are real and lengthy expressions of the $\chi_{ijk}$. They can be considered as independent fit parameters, as the susceptibility contains $36$ independent real elements for $1$ symmetry. We have obtained fits both in terms of the ten fit parameters $a_i$ and $b_i$ as well as in terms of the $\chi_{ijk}$, but we state only the values for  $a_i$ and $b_i$ obtained from best fits in Table~\ref{tab:1_fit_values}. The table also contains $R^2$ values, which are all found to be $0.99$. The resulting fits are shown in panels (b, f) and (c, g) of Fig.~1 in the main text. 

  For completeness, Table~\ref{tab:outgoing_I_others} shows the general form of the outgoing intensities $I^{[1\bar{1}0]}_{\text{SHG}}$ and $I^{[11\bar{1}]}_{\text{SHG}}$ for all time-reversal invariant (i.e., grey) subgroups of 4mm1' along with the corresponding crystallographic groups, where time-reversal symmetry is broken. It is interesting to note that as long as the diagonal mirror symmetry $m_{x,x,z}$ is present, the form of the outgoing intensities is identical to the fully symmetric case with spatial 4mm symmetry.  
  \onecolumngrid
  \begin{table}[h!]
  \begin{center}
  \begin{tabular}{c||c|c}
  1 & $\Delta t = 0.0$~ps & $\Delta t = 1.0$~ps
\\
  \hline 
  \hline
  $a_1/a_5$ & $0.078$& $0.078$ \\
  $a_2/a_5$ & $0.15$& $0.15$ \\
  $a_3/a_5$ & $-0.031$& $-0.031$ \\
  $a_4/a_5$ & $0.087$& $0.086$ \\
  \hline
  $\mathcal{C}^{[11\bar{1}]}_0/a_5 = b_1/a_5$ & $20$& $29$ \\
  $\mathcal{C}^{[11\bar{1}]}_1/a_5 =b_2/a_5$ & $3.2$ & $2.6$ \\
  $\mathcal{C}^{[11\bar{1}]}_3/a_5 =b_3/a_5$ & $2.2$& $1.9$ \\
  $\mathcal{C}^{[11\bar{1}]}_4/a_5 =b_4/a_5$ & $1.7$& $0.70$ \\
  $\frac12 \mathcal{C}^{[11\bar{1}]}_2/a_5 =b_5/a_5$ & $-0.45$& $-0.29$ \\
  \hline \hline
  $R^2([1, \bar{1},0])$ & $0.99$& $ 0.99$ \\ 
  $R^2([1,1,\bar{1}])$ & $0.99$ & $0.99$  \\
  \hline
% \hline
% \multicolumn{2}{c}{Item} \\
% \cline{1-2}
  \end{tabular}
  \caption{Fit parameters and corresponding $R^2$ values for a $1$ point group symmetry of the photoexcited state. The general form of the outgoing intensities is (see Table~\ref{tab:outgoing_I_relevant}): $I^{[1\Bar{1}0]}_{\text{SHG}} = a_1 c_\phi^4 + a_2 c_\phi^3 s_\phi + a_3 c_\phi s_\phi^3 + a_4 s_\phi^4 + a_5 s_{2 \phi}^2$ and $I^{[11\Bar{1}]}_{\text{SHG}} = b_1 c_\phi^4 + b_2 c_\phi^3 s_\phi + b_3 c_\phi s_\phi^3 + b_4 s_\phi^4 + b_5 s_{2\phi}^2$. We note that in the main text, we use the following notation for $1$ symmetry, $I_{\text{SHG}}^{[11\bar{1}]} = \sum_{n=0}^4 \mathcal{C}_n^{[11\bar{1}]} \sin^n(\phi) \cos^{4-n}(\phi)$, corresponding to $\mathcal{C}^{[11\Bar{1}]}_0 = b_1$, $\mathcal{C}^{[11\Bar{1}]}_1 = b_2$, $\mathcal{C}^{[11\Bar{1}]}_2 = 2 b_5$, $\mathcal{C}^{[11\Bar{1}]}_3 = b_3$, and $\mathcal{C}^{[11\Bar{1}]}_4 = b_4$. The overall rotation of the $I^{[11\bar{1}]}_{\text{SHG}}$ pattern and anisotropy in the photoinduced lobes are captured by the $b_2$ coefficient. Asymmetry in the photoinduced lobes is given by a non-zero $b_3, b_4$ coefficients (predominately $b_3$ in our fits). The coefficient $b_4$ is responsible for a finite value of the local minima around $\phi = 90^{\circ}, 270^{\circ}$.}
  \label{tab:1_fit_values}
  \end{center}
  \end{table}
  \onecolumngrid

  \begin{table}[h!]
  \begin{center}
  \begin{tabular}{c||c|c}
  MPG & Form of $I_\text{SHG}^{\text{[1$\bar{1}$0]}}$ & Form of $I_\text{SHG}^{\text{[11$\bar{1}$]}}$ 
\\
  \hline 
  \hline
  2mm1', $(m_{x,-x,z}, m_{x,x,z})$&  $a_1 s_{2 \phi}^2$ & $\bigl(b_1 + b_2 c_\phi^2  \bigr)^2$ \\ 
  2mm, $(m_{x,-x,z}, m_{x,x,z})$ & $a_1 s_{2 \phi}^2$ & $b_1 c_\phi^4 + b_2 s_{\phi}^4 +b_3 s_{2\phi}^2$\\
  \hline
  2mm1', $(m_{x,0,z}, m_{0,y,z})$&  $\bigl(  a_1 c_\phi^2 + a_2 s_{2\phi}  \bigr)^2$ & $\bigl(b_1  + b_2 c_\phi^2 + b_3 s_{2\phi}   \bigr)^2$\\ 
  2mm, $(m_{x,0,z}, m_{0,y,z})$ & $c_\phi^2 \bigl(a_1 + a_2 c_\phi^2 + a_3 s_{2 \phi} \bigr)$ & $b_1 c_\phi^4 + b_2 c_\phi^3 s_\phi + b_3 c_\phi s_\phi^3 + b_4 s_\phi^4 + b_5 s_{2\phi}^2$\\
  \hline
  41'&  $\bigl(a_1 c^2_\phi + a_2 s_{2\phi} \bigr)^2$ & 
  $\bigl(b_1 + b_2 c_\phi^2+ b_3 s_{2\phi} \bigr)^2$ \\ 
  4& $c_\phi^2 \bigl(a_1 + a_2 c_\phi^2 + a_3 s_{2 \phi} \bigr)$ & $b_1 c_\phi^4 + b_2 c_\phi^3 s_\phi + b_3 c_\phi s_\phi^3 + b_4 s_\phi^4 + b_5 s_{2\phi}^2$\\
  \hline
  m1' $(m_{x,0,z})$&  $\bigl(a_1 + a_2 c_\phi^2 + a_3 s_{2 \phi}\bigr)^2$ & $\bigl(b_1 + b_2 c_\phi^2 + b_3 s_{2\phi} \bigr)^2$ \\
  m $(m_{x,0,z})$  &  $a_1 c_\phi^4 + a_2 c_\phi^3 s_\phi + a_3 c_\phi s_\phi^3 + a_4 s_\phi^4 + a_5 s_{2 \phi}^2$& $b_1 c_\phi^4 + b_2 c_\phi^3 s_\phi + b_3 c_\phi s_\phi^3 + b_4 s_\phi^4 + b_5 s_{2\phi}^2$ \\
  \hline
  21' &  $\bigl(  a_1 c_\phi^2 + a_2 s_{2\phi}  \bigr)^2$ & $\bigl(b_1 + b_2 c_\phi^2 + b_3 s_{2\phi} \bigr)^2$ \\
  2 & $c_\phi^2 \bigl(a_1 + a_2 c_\phi^2 + a_3 s_{2 \phi} \bigr)$ & $b_1 c_\phi^4 + b_2 c_\phi^3 s_\phi + b_3 c_\phi s_\phi^3 + b_4 s_\phi^4 + b_5 s_{2\phi}^2$ \\
  \hline
% \hline
% \multicolumn{2}{c}{Item} \\
% \cline{1-2}
  \end{tabular}
  \caption{General form of the outgoing intensities along $[1\bar{1}0]$ (s-out) and $[11\bar{1}]$ (p-out) for the remaining white and corresponding gray MPGs that are derived from 4mm1'. Here, $c_\phi \equiv \cos \phi$, $s_\phi \equiv \sin \phi$ and the coefficients $a_i$ and $b_i$ are real. The expression for $m1'$ and $m$ with mirror $m_{0,y,z}$ is identical to the one with $m_{x,0,z}$. We state these expressions for completeness, but note that we do not consider other possible black-white magnetic subgroups of 4mm1', as they are not relevant to our experiment. }\label{tab:outgoing_I_others}
\end{center}
\end{table}

%\newpage
 \onecolumngrid

\subsection{Discussion of symmetry breaking transient features in the SHG patterns}
\label{subsec:discussion_fit}
  Let us briefly discuss which of the terms in the general form of the outgoing intensities allow us to capture the observed transient features in the SHG pattern. To recall, in the transient regime at $\Delta t = 0.0$~ps and $\Delta t = 1.0$~ps, we find (i) an overall rotation of the  $I^{[11\bar{1}]}_{\text{SHG}}$ pattern by $\sim2.5^{\circ}$, and (ii) emergent, asymmetric lobes at $\phi = 90^{\circ}$ and $\phi = 270^{\circ}$. In contrast the $I^{[1\bar{1}0]}_{\text{SHG}}$ pattern remains unchanged under pump excitation.

  Importantly, neither of the two features, (i) and (ii), described above can be captured by a tensor constrained by 4mm1' symmetry, as the lobes must remain pinned to the coordinate axes. While the emergence of small lobes at $\phi = 90^{\circ}$ (and $\phi = 270^{\circ}$) can be enforced by increasing the value of $b_1$, these will necessarily be symmetric around a maximum at $\phi = 90^{\circ}$. Similarly, the overall rotation can be accounted for in the absence of time-reversal symmetry for 4mm, but the asymmetry of the photoinduced lobes at $\phi = 90^{\circ}$ and $\phi = 270^{\circ}$ cannot be obtained with a 4mm tensor. Since the diagonal mirror $m_{x,x,z}$ enforces the form for m1' (m) to be identical to 4mm1' (4mm) (see Table~\ref{tab:outgoing_I_relevant}), the same applies for a (surface) tensor constrained by m1' (with $m_{x,x,z}$ mirror symmetry). 
  
  Interestingly, the asymmetry of the small lobes at $\phi = 90^{\circ}$ and $\phi = 270^{\circ}$ cannot be produced in the presence of time-reversal symmetry, even if all spatial symmetries are broken, i.e. for 1' symmetry. This is shown most transparently by rewriting 
\begin{equation}
    (b_1 \sin^2 \phi + b_2 \cos^2\phi + b_3 \sin \phi \cos \phi)^2 = \bigl[a_1 \cos^2 (\phi - \phi_0) + a_2 \sin^2(\phi - \phi_0)\bigr]^2
\end{equation}
with global shift angle $\phi_0 = \frac12 \sin^{-1} [b_3/(a_1 - a_2)]$ and $a_1 = \frac{b}{2} \mp \frac{\sqrt{\beta}}{2}$, $a_2 = \frac{b}{2} \pm \frac{\sqrt{\beta}}{2}$ , where $b = b_1 + b_2$ and $\beta = (b_1 - b_2)^2 + b_3^2$. The sign in the expressions for $a_1$ and $a_2$ is chosen such that $\text{sign} (a_1 - a_2) = \text{sign} (b_1 - b_2)$. Note that $a_1 - a_2 = \mp \sqrt{\beta}$. While the expression for 1' can thus reproduce a global shift of the pattern by $\phi_0$, the pattern is necessarily symmetric around the lobes and in particular the small side lobes close to $\pi/2$. In contrast, the observed asymmetry shown in Fig.~1 (b, f, c, g) is fully consistent with 1 symmetry. Hence, the asymmetry of these emergent lobes can be directly associated with a breaking of both time-reversal and mirror $m_{x,x,z}$ symmetry brought on by photocurrent generation. 

In our fit using 1 symmetry, the overall rotation is (mostly) accounted for by the coefficient $b_2 \equiv \mathcal{C}^{[11\bar{1}]}_1$ (see Table~\ref{tab:1_fit_values}). The asymmetry of the photoinduced lobes at $\phi = 90^{\circ}$ and $\phi = 270^{\circ}$ is (mostly) expressed by the fit parameter $b_3\equiv \mathcal{C}^{[11\bar{1}]}_3$, because it is multiplied by $\sin^3 \phi \cos\phi$ and is thus larger close to $\phi = \pi/2$ than $\sin \phi \cos^3 \phi$. The magnitude of the side lobes at $\phi = \pi/2$ is encoded by the fit parameter $b_4\equiv \mathcal{C}^{[11\bar{1}]}_4$ (which is multiplied by $\sin^4 \phi$). While the largest parameter is $b_1\equiv \mathcal{C}^{[11\bar{1}]}_0$, which is responsible for the main lobes at $\phi = 0^{\circ}$, we find that at $\Delta t = 0.0 (1.0)$~ps the "overall rotation parameter" $b_2/b_1 = \mathcal{C}^{[11\bar{1}]}_1/ \mathcal{C}^{[11\bar{1}]}_0= 0.16 (0.09)$ and ``lobe asymmetry parameter'' $b_3/b_1 = \mathcal{C}^{[11\bar{1}]}_3/ \mathcal{C}^{[11\bar{1}]}_0 = 0.11 (0.07)$ are still significant. In other words, a symmetry breaking photocurrent has a significant impact of order 10 - 15\% on the $I^{[11\bar{1}]}_{\text{SHG}}$ pattern.

While it is evident that photocurrent generation leads to clear changes in the TR-SHG pattern as compared to equilibrium, there remains an open question as to whether transient photocurrents can break symmetries of the electronic wavefunctions over a wide energy window, or simply open an additional nonlinear optical channel beyond that of equilibrium SHG. In both scenarios, a transient photocurrent is the underlying mechanism, and this is nontrivial, as both describe the modification of an intrinsic material property over picosecond timescales following pump excitation. Since both cases describe a photocurrent-induced change of the second harmonic response tensor, $\chi^{(2)}$, that is proportional to the photocurrent, we cannot unambiguously distinguish between them in our experiment. However, consideration of the photocurrent decay may allow for these two scenarios to be distinguished, as a direct change in the electronic wavefunction will be proportional to the photocurrent itself, while the opening of a new nonlinear optical channel in the form of $E(2\omega)\propto E_{Photocurrent}(\omega_{THz})E(\omega)E(\omega)$ is proportional to the time derivative of the photocurrent and thus entails a decay time. Addressing which scenario dominates in the TR-SHG response is beyond the scope of this work, but it is worth mentioning that our observation of transient symmetry breaking only when the probe is resonant with the pump implies that a deformation of the carrier distribution from equilibrium plays an important role. Importantly, this does not exclude a significant impact on low-energy Weyl node carriers, as these carriers can couple during the relaxation process via Coulomb interaction and scattering, but further study is needed to determine how strong a coupling exists between optical photocurrents and low-energy carriers.   

\newpage
\section{Consideration of Alternative Electronic Origins for the Transient Symmetry Broken State}

Our analysis in X. systematically considers every relevant sub-group of $4mm1'$, including contributions from surface SHG, and demonstrates both time-reversal and diagonal mirror ($m_{xxz}$) symmetry must be lifted in order to capture the degree of symmetry breaking observed in our experiments. Together with the polarization dependence shown in Fig. 4 of the main text, where control over the degree of symmetry breaking in the SHG pattern is illustrated, these conditions place several constraints on any alternative electronic origin for the transient symmetry-broken state. Ultrafast x-ray diffraction, shown in III, allows us to rule out any structural dynamics occurring on the sub-ps timescale of the reduced symmetry state, including coherent phonon generation, which is likewise absent from our TR-SHG data. Furthermore, Figs. \ref{fig:Fluence_Dep}(b) and \ref{fig:Temp_Dep}(b) reveal insignificant variation in the dynamics of the asymmetric, photoinduced lobe with fluence or temperature, ruling out these factors even during the initial non-equilibrium stages after photoexcitation. With these additional constraints, we focus below on alternatives that either introduce directional anisotropy or could give rise to transient, polarization-dependent, time-reversal symmetry breaking.    

\begin{figure}[b]
    \centering     
    \includegraphics[width=\columnwidth]{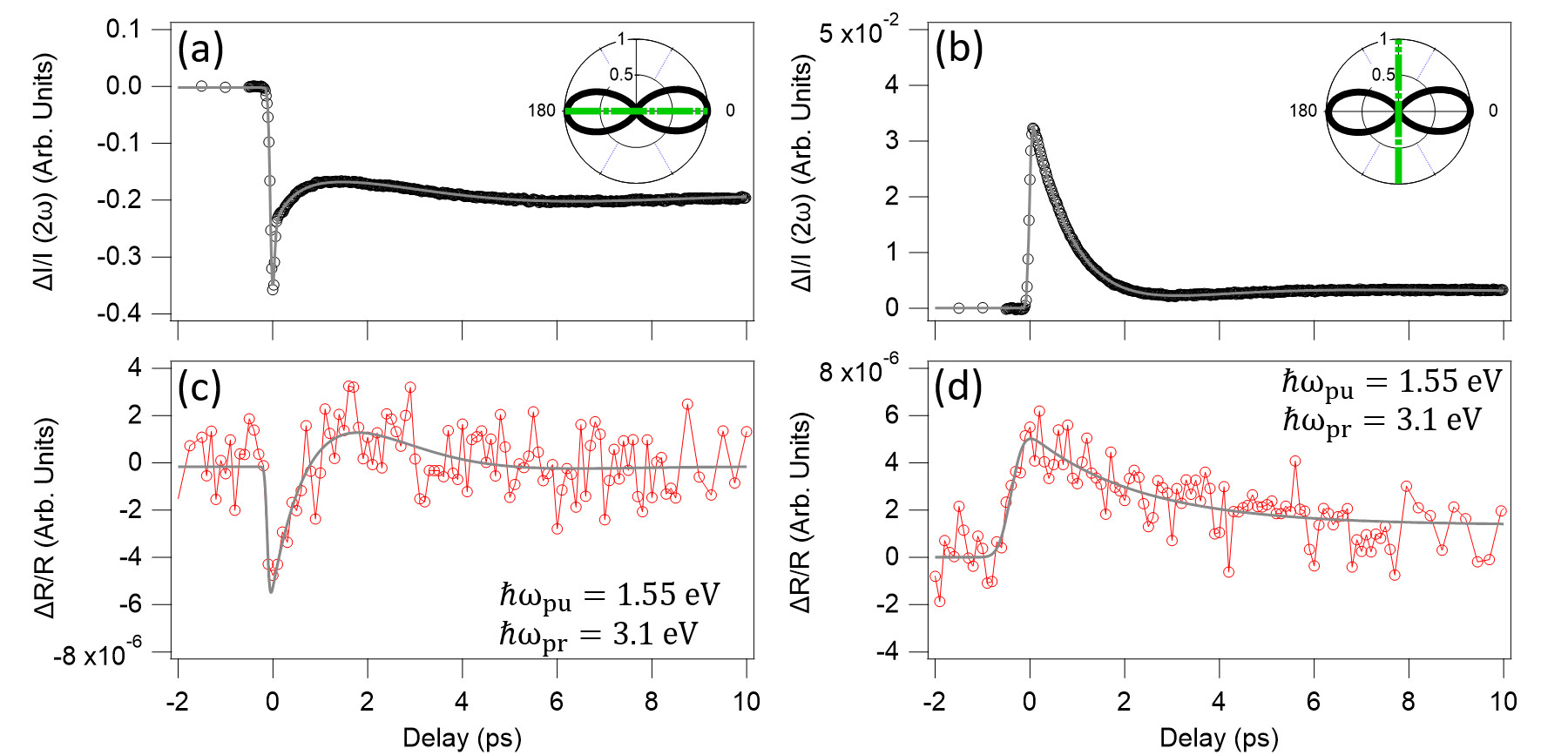}
    \caption{Comparison of TR-SHG traces measured for input pump polarizations along the (a) [1,1,$\Bar{1}$] and (b) [1,$\Bar{1}$,0] axes (green dashed line) of the [1,1,$\Bar{1}$] SHG pattern (insets) obtained for $\hbar\omega_{pu}= 1.55$ eV with non-degenerate optical pump-probe traces ($\hbar\omega_{pu} = 1.55$ eV; $\hbar\omega_{pr} = 3.1$ eV) taken with an input pump polarization parallel to the (c) [1,1,$\Bar{1}$] and (d) [1,$\Bar{1}$,0] axes. Here a pump fluence of 4.34 mJ/cm$^{2}$ was used for both TR-SHG and non-degenerate optical pump-probe experiments.}
    \label{fig:Anisotropy}
\end{figure}

\underline{Spatially inhomogeneous pump volume:} Our pump beam diameter (65 $\mu$m) is $1.4\times$ larger than our probe beam diameter (45 $\mu$m), ensuring that we initially probe an approximately homogeneous distribution of photoexcited carriers. The observed transient symmetry changes in our TR-SHG signal near $90^{\circ}$ and $270^{\circ}$ are short lived ($<1$ ps), so any spatial inhomogeneities would have to develop on the same timescale, much faster than carriers can diffuse in TaAs \cite{Weber_MPTP_2017}. 

\underline{Sum frequency generation:} Transient symmetry breaking lasts on a timescale $\sim 10\times$ longer than the temporal overlap of the $\sim 80$ fs pump and probe pulses. While a degenerate TR-SHG experiment is necessary for measuring a photocurrent-induced transient symmetry change, we can discount any contribution arising from sum frequency generation due to this large discrepancy in timescale between the reduced symmetry state and the temporal overlap of the pump and probe pulses.

\underline{Anisotropic changes in optical constants:} Fig. \ref{fig:Anisotropy} compares TR-SHG traces measured for input pump polarizations along the (a) [1,1,$\Bar{1}$] and (b) [1,$\Bar{1}$,0] axes with non-degenerate transient reflectivity taken with a 1.55 eV optical pump pulse polarized along these same directions. Here, photoinduced modulation of the reflectivity, probed at 3.1 eV, reveals that light-induced changes to optical constants, using the same excitation (1.55 eV) energy as in our TR-SHG experiments, are 4 - 5 orders of magnitude smaller than the TR-SHG response in both directions. This discrepancy implies that our observation of transient symmetry breaking within the SHG pattern cannot be attributed to photoinduced anisotropy of linear optical constants. 

\underline{Anisotropic carrier thermalization:} In graphene, an initial anisotropic electron distribution, linked to pseudospin flipping from interband transitions \cite{Trushin_EPL_2011}, can be generated by pumping with linearly polarized light \cite{Malic_APL_2012}. Due to the predominance of collinear Coulomb scattering, this anisotropy can persist, but is ultimately lost with increasing pump fluence through the emergence of a noncollinear scattering mechanism, generating an isotropic thermalization distribution \cite{Noncollinear_PhysRevLett.117.087401}. Such behavior contrasts with what is observed in Fig. \ref{fig:Fluence_Dep}(b), where the emergent asymmetric, photoinduced lobe present in the [1,1,$\Bar{1}$] SHG pattern increases linearly with pump fluence. This suggests the emergent elements seen in the transient pattern are proportional to the strength of the photocurrent, which itself is linearly dependent on fluence or intensity. Thus, the degree of symmetry breaking is found to scale with intensity as expected for a current-induced SHG mechanism \cite{Khurgin_1995}.  

\underline{Polarization dependent magneto-optical response:} One last possibility is the inverse Faraday effect, which effectively applies a transient magnetic field along the Poynting vector of a circularly polarized light pulse, lasting over the course of the pulse duration \cite{IFE_RevModPhys.82.2731}. However, as shown in Fig. 3(b) of the main text, a photoinduced symmetry-breaking feature is present under both circular and linearly polarized excitation in our experiments, and our ability to control symmetry in the [1,1,$\Bar{1}$] pattern under linearly polarized excitation in Fig. 4(d) of the main text effectively rules out this possibility.   

\newpage
\section {First-principles calculations of optical conductivity in TaAs}

\begin{figure}[h]
    \includegraphics[width=\linewidth]{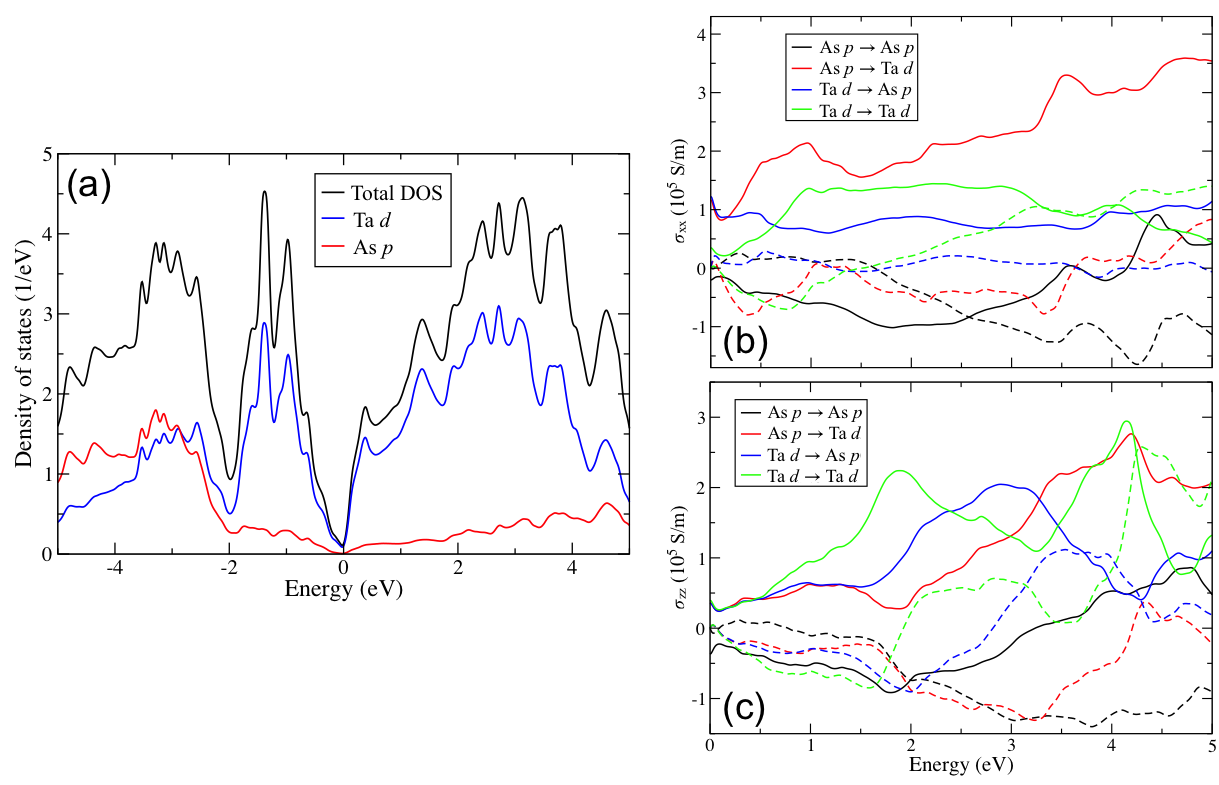}
    \caption{ (a) Total density-of-states (DOS) of TaAs (black) decomposed into Ta-d (blue) and As-p (red) orbital contributions. Calculated optical conductivity for (b) $\sigma_{xx}=\sigma_{yy}$ and (c) $\sigma_{zz}$ decomposed to show different orbital contributions. Here, solid and dashed curves represent the real and imaginary parts of the optical conductivity, respectively.}
    \label{fig_S8:abinitio}
  \end{figure}

  The calculated density-of-states (DOS) of TaAs is shown in Fig.~\ref{fig_S8:abinitio}(a). The DOS near the Fermi energy (0 eV) mainly comes from the contribution of Ta-d orbitals. From the DOS, we find that the optical conductivity can be described by four contributions: the transitions from As-p to As-p orbitals, As-p to Ta-d orbitals, Ta-d to As-p orbitals, and Ta-d to Ta-d orbitals. In the Kubo-Greenwood formula, there are two momentum matrix elements in the numerator. The orbital contribution for the transitions can be defined via one of the momentum matrices through, $\langle kM|\hat{p}_{x~\lor~ y~\lor~z}|kN\rangle$, where M and N denote the pseudoatomic orbitals. The orbital contribution for the optical conductivity, $\sigma(\omega)$ is shown in Fig.~\ref{fig_S8:abinitio}(b-c) for the $\sigma_{xx}$ and $\sigma_{zz}$ components, respectively. In Fig.~\ref{fig_S8:abinitio}(b), a prominent contribution close to $1.0$ eV in the in-plane conductivity $\sigma_{xx}$ (or $\sigma_{yy}$) can be identified as the result of an As-p to Ta-d transition. In contrast, Fig.~\ref{fig_S8:abinitio}(c) shows a prominent contribution close to $1.9$ eV for the out-of-plane conductivity $\sigma_{zz}$, which can be identified as the result of a Ta-d to Ta-d transition. 
%\end{center}
\newpage
\bibliography{Sirica_Main_v1.bib}
\end{document}